# Skein Construction of Balanced Tensor Products

Manuel Araújo [*], Jin-Cheng Guu [†], Skyler Hudson [‡]

**Abstract**

The theory of tensor categories has found applications across various fields, including representation theory [Eti+15], quantum field theory (conformal in 2 dimensions [Fje+06] [Cre+24], and topological in 3 and 4 dimensions), quantum invariants of low-dimensional objects [Tur10] [Bel+23], topological phases of matter [LW05] [KWZ17] [NRW23], and topological quantum computation [Fre+03]. In essence, it is a categorification of the classical theory of algebras and modules. In this analogy, the Deligne tensor product $\boxtimes$ is to the linear tensor $\otimes_{\mathbb{C}}$ as the balanced tensor product $\boxtimes_{\mathrm{C}}$ is to the tensor over algebra $\otimes_{A}$, where $\mathbb{C}$ is a field, $A$ is a $\mathbb{C}$-algebra, and $\mathrm{C}$ is a tensor category.

Before this work, several algebraic constructions for balanced tensor products were known, including categories of modules [DSS19], internal Hom spaces [DN12], and generalized categorical centers [Eti+09] [JT22] [Hoe19]. In this paper, we introduce a topological construction based on skein theory that offers a better mix of algebra and topology. This approach not only works for products of multiple module categories, but also provides the missing key to proving that the Turaev-Viro state sum model [TV92] naturally arises from the 3-functor in the classification of fully extended field theories [Lur09]. Building on this result, we establish this long-anticipated proof in an upcoming work [GAH25].

Compiled Time: [2025-01-09 22:54:29]

# Contents

[*] <manuel.araujo@tecnico.ulisboa.pt>
[†] <guu@ualberta.ca>, University of Alberta
[‡] <smhudson@umass.edu>, University of Massachusetts Amherst

# Acknowledgement

The authors express their gratitude to Thibault Décoppet, Terry Gannon, Owen Gwilliam, Ignacio López Franco, Christoph Schweigert, Ying Hong Tham, and Harshit Yadav for their feedback and discussions.

One of the authors, J-C Guu, would like to express appreciation for the support provided by UMass Amherst, U of Alberta, and the Pacific Institute for the Mathematical Sciences (PIMS) during the completion of this work. Therefore the research conducted for this paper is supported by the Pacific Institute for the Mathematical Sciences (PIMS). The research and findings may not reflect those of the Institute.

# 1 Introduction

The notion of a balanced tensor product was first given in [Eti+09], and mainly developed in [DSS19] and [DSS21]. It has different interpretations in various fields:

- In the theory of topological phases, this construction corresponds to fusing two domains (labeled by module categories) along the domain wall (labeled by tensor categories) [KWZ17].
- In the theory of extended topological quantum field theory, it is the 1-categorical structure of a common target 3-category TC [DSS21].
- In the theory of vertex operator algebras (VOA), this construction helps classify the full bulk CFTs (as explained in [Gan23]), and constructing new VOAs [EG24].
- In the theory of tensor categories itself, it closely relates to the categorical center construction. In particular, the Drinfeld center $Z(C)$ of $C$ is equivalent to $C \boxtimes_{C \boxtimes C^{op}} C$ [Jr11] [DSS21].

Before this work, several algebraic constructions for balanced tensor products were known, including categories of modules [DSS19], internal Hom spaces [DN12], and generalized categorical centers [Eti+09] [JT22] [Hoe19]. These approaches well-captured the algebraic aspects of balanced tensor products.

In this paper, we provide a topological construction based on skein theory, which strikes a better mix between algebra and topology. It has a few advantages:

1. Its topological nature allows us to use skein diagrams to calculate the values of the fully extended 3D field theories associated to a fusion category by the cobordism hypothesis [Lur09] [DSS21] [GAH25].
2. It generalizes seamlessly to balanced tensor products involving more than two module categories: $M_1 \boxtimes_{C_1} M_2 \boxtimes_{C_2} M_3 \dots$. (see the picture in 2.34).

Some of the (potential) applications of this work are as follows.

1. **(Skein Nature of eTFTs)** In an upcoming work [GAH25], we build on the main result presented here to prove the long-anticipated theorem that the Turaev-Viro state sum model [TV92] naturally and necessarily emerges from the 3-functor in the classification of fully extended field theories [Lur09] [DSS21].
2. **(No-Go: Detection of Exotic Smooth Structures)** We anticipate that this approach can be extended to a 4-dimensional analogue of Turaev-Viro theory, the Crane-Yetter model. Additionally, a similar strategy should apply to a broader generalization based on fusion 2-categories, as explored in [DR18]. Proving this would lend support to the conjecture proposed there, which suggests that despite the construction's generality and power, it remains an extended topological field theory and therefore cannot detect exotic structures [Reu20].
3. **(Factorization Homology)** The main result of this paper provides a simple reproof of the result in [JT22] that shows equivalence between the Crane-Yetter model and factorization homology [AFR15] in 4D (the excision property).
4. **(Defected TFTs)** We expect work being useful in the study of defect field theories, such as in [Meu22].

5. **(Computations in Tensor / Higher Categories)** Given that the Drinfeld center is a specific instance of the balanced tensor product, we expect this paper will contribute to recent advancements, particularly the explicit computation of centers as explored in [MT24], and further develop the computational aspects of balanced tensor products.

## 1.1 Outline

Given a finite semisimple tensor category $C$, and finite semisimple module categories $M_C$ and ${}_C N$ (over $C$), we aim to provide a mathematical construction of their balanced tensor product $M \boxtimes_C N$ based on skein theory. We start by defining the preskein category $M \otimes_C N$ (2.1), which satisfies a similar defining universal property (2.14), but fails to be $k$-linear. We then (Cauchy) complete it to the skein category $sk(M, C, N)$ (1.10), and prove that it is indeed equivalent to $M \boxtimes_C N$ when $k$ has characteristic $0$ (Main Theorem, 2.33). We also obtain analogous results in prime characteristic (see 2.31 and 2.32). This generalizes immediately to the balanced tensor product of multiple module categories: $M_1 \boxtimes_{C_1} M_2 \boxtimes_{C_2} M_3 \dots$ (2.35). We then drop the semisimplicity assumption, and prove a similar, yet more general, statement (A.30), where the balanced tensor product is generalized to Kelly balanced tensor product.

## 1.2 Preliminaries

Throughout the whole paper, unless mentioned otherwise, we fix a field $k$ and denote by $\mathrm{Vect}_k$ the category of $k$-vector spaces. We assume all categories, functors and natural transformations are enriched over $\mathrm{Vect}_k$.

**Definition 1.1** (Linearized Product, $M \otimes N$)
Given categories $M, N$ we denote by $M \otimes N$ the category whose objects are pairs $(m, n)$ with $m \in M$ and $n \in N$ and where $\mathrm{Hom}_{M \otimes N}((m, n), (m', n')) = \mathrm{Hom}_M(m, m') \otimes \mathrm{Hom}(n, n')$. Heuristically, $M \otimes N$ is $M \times N$ with bi-linearlized hom spaces. ⋄

**Remark 1.2** (Categorical Products: $M \otimes N$, $M \otimes_C N$, $M \hat{\otimes}_C N$, $M \boxtimes N$, $M \boxtimes_C N$)
For convenience, we collect the six kinds of categorical products we use throughout the paper.

1. Linearized product $M \otimes N$ (1.1).
2. Pre-skein category $M \otimes_C N$ (2.1).
3. Deligne tensor product $M \boxtimes N$ (1.11).
4. Balanced tensor product $M \boxtimes_C N$ (1.10).
5. General Pre-skein category $M \hat{\otimes}_C N$ (A.14).
6. Kelly Balanced tensor product $M \boxtimes_C^K N$ (A.21).

⋄

**Definition 1.3** ((Bi)linear functor, $\mathrm{Fun}(M, N)$, $\mathrm{Lin}(M, N)$, $\mathrm{Bilin}(M \otimes N, L)$)
Given categories $M, N, L$, denote by $\mathrm{Fun}(M, N)$ the category of ($\mathrm{Vect}_k$-enriched) functors from $M$ to $N$. A functor $F \in \mathrm{Fun}(M, N)$ is linear if it is right exact (i.e. if it preserves all finite colimits). We denote the category of linear functors $M \to N$ by $\mathrm{Lin}(M, N)$. A functor $F \in \mathrm{Fun}(M \otimes N, L)$ is called bilinear if it is linear in each variable. We denote the category of bilinear functors $M \otimes N \to L$ by $\mathrm{Bilin}(M \otimes N, L)$. ⋄

**Definition 1.4** ($k$-linear category) [DSS19]
A $k$-linear category is an abelian category with a compatible $\mathrm{Vect}_k$-enrichment. ⋄

**Definition 1.5** [Eti+15] An object in a $k$-linear category is simple if it has no nontrivial subobjects. A $k$-linear category is semisimple if every object splits as a direct sum of simple objects. A $k$-linear category $M$ is finite if

- $\mathrm{Hom}_M(X, Y)$ is finite dimensional, for all objects $X, Y \in M$.
- Every object of $M$ has finite length.

- $M$ has enough projectives.
- $M$ has finitely many isomorphism classes of simple objects.

A k-linear monoidal category is a k-linear category $C$ with a monoidal structure where the functor $C \otimes C \to C$ is bilinear. A tensor category is a rigid k-linear monoidal category. $\diamond$

In a finite semisimple k-linear category, every object splits as a finite direct sum of simple objects. When $M, N$ are semisimple, every $\mathrm{Vect}_k$-enriched functor $M \to L$ is automatically exact (thus linear) and every $\mathrm{Vect}_k$-enriched functor $M \otimes N \to L$ is automatically bilinear.

**Assumption 1.6** When $C$ is a k-linear monoidal category (e.g. a tensor category), any $C$-module category $M$ is assumed to be k-linear and the structure map $C \otimes M \to M$ is assumed to be bilinear.

Given a right (left, resp.) $C$-module category $M_C$ (${}_C N$, resp.), we denote by $m \triangleleft c$ ($n \triangleright n$, resp.) the object in $M$ ($N$, resp.) that results from acting with $c \in C$ on $m \in M$. $\diamond$

## 1.3 Balanced Tensor Product

In this section we recall the definition of the balanced tensor product from [DSS19].

**Definition 1.7** (Balanced Functor)
Let $C$ be a monoidal category, $M_C$, ${}_C N$ $C$-module categories, and $L$ a category. Then a $C$-balanced functor $M \otimes N \to L$ is a pair

$$(M \otimes N \xrightarrow{F} L, \alpha)$$

where $F$ is a ($\mathrm{Vect}_k$-enriched) functor and $\alpha$ is a natural isomorphism

$$\begin{array}{ccccc} M \otimes C \otimes N & \longrightarrow & M \otimes N & \xrightarrow{F} & L \\ & \searrow & \Downarrow \alpha & \nearrow_{F} & \\ & & M \otimes N & & \end{array}$$

where the object $(m, c, n)$ in $M \otimes C \otimes N$ is sent by the first upper arrow to $(m \triangleleft c, n)$, and by the first lower arrow to $(m, c \triangleright n)$. $\diamond$

**Definition 1.8** (Balanced Natural Transformation)
Let $C$ be a monoidal category, let $M_C$, ${}_C N$ be $C$-module categories and let $L$ be a category. Let $(F, \alpha), (G, \beta)$ be $C$-balanced functors from $M \otimes N$ to $L$. A $C$-balanced natural transformation from $(F, \alpha)$ to $(G, \beta)$ is a natural transformation $F \xrightarrow{\eta} G$ such that the following diagram commutes.

$$\begin{array}{ccc} F(m \triangleleft c, n) & \xrightarrow{\alpha_{m,c,n}} & F(m, c \triangleright n) \\ {\scriptstyle \eta_{m \triangleleft c, n}} \downarrow & & \downarrow {\scriptstyle \eta_{m, c \triangleright n}} \\ G(m \triangleleft c, n) & \xrightarrow[\beta_{m,c,n}]{} & G(m, c \triangleright n) \end{array}$$

$\diamond$

**Definition 1.9** ($\mathrm{Fun}^{C-bal}(M \otimes N, L)$ and $\mathrm{Bilin}^{C-bal}(M \otimes N, L)$)
Let $C$ be a monoidal category, let $M_C$, ${}_C N$ be $C$-module categories and let $L$ be a category. Then the category of $C$-balanced functors $\mathrm{Fun}^{C\text{-bal}}(M \otimes N, L)$ is defined to be the category whose objects are $C$-balanced functors $M \otimes N \to L$ and whose morphisms are $C$-balanced natural transformations.

We denote by $\mathrm{Bilin}^{C\text{-bal}}(M \otimes N, L)$ the full subcategory of $\mathrm{Fun}^{C\text{-bal}}(M \otimes N, L)$ whose objects are bilinear (thus right exact) $C$-balanced functors. $\diamond$

**Definition 1.10** (Balanced Tensor Product)
Let $C$ be a k-linear monoidal category, and let $M_C$, ${}_CN$ be k-linear module categories over $C$. Then the balanced tensor product $M \boxtimes_C N$ is a k-linear category together with a $C$-balanced bilinear functor $M \otimes N \to M \boxtimes_C N$ such that

$$\mathrm{Lin}(M \boxtimes_C N, L) \to \mathrm{Bilin}^{C-bal}(M \otimes N, L)$$

is an equivalence, for any k-linear category $L$. ◇

In [DSS19] the authors construct $M \boxtimes_C N$ for any finite tensor category $C$ and any finite $C$-module categories $M, N$, thus establishing existence in this setting. They do this by using the fact that the module categories $M, N$ can be realized as categories of modules over algebra objects $A_M$ and $A_N$ in $C$ and then showing that the category of $A_M - A_N$-bimodule objects in $C$ is k-linear and satisfies the required universal property. In the present paper, we will give an alternative construction of $M \boxtimes_C N$ when $M$, $N$, and $M \boxtimes_C N$ are semisimple.

**Remark 1.11** Notice that the Deligne tensor product $M \boxtimes N$ is just $M \boxtimes_{\mathrm{Vect}_k} N$. ◇

# 2 Skein Construction

## 2.1 Pre-skein Category $M \otimes_C N$

In this section, we present the pre-skein category $M \otimes_C N$ and develop some of its properties.

**Definition 2.1** (Pre-Skein Category) For $C$ a monoidal category having right duals and $M_C$, ${}_CN$ module categories, we define the pre-skein category $M \otimes_C N$ as follows. Its objects are pairs $(m, n)$ with $m \in M$ and $n \in N$, which we denote by $m \boxtimes n$. The hom space $\mathrm{Hom}_{M \otimes_C N}(m \boxtimes n, m' \boxtimes n')$, as a k-vector space, is the quotient of the k-vector space

$$V((m,n),(m',n')) := \bigoplus_{c,\bar{c} \in \mathrm{Obj}(C)} \mathrm{Hom}_M(m, m' \triangleleft c) \otimes \mathrm{Hom}_C(c, \bar{c}) \otimes \mathrm{Hom}_N(\bar{c} \triangleright n, n')$$

by the k-linear subspace spanned by

$$\phi \otimes (\bar{\pi} \circ \pi) \otimes \psi - (\phi_\pi \otimes \bar{\pi} \otimes \psi) \tag{2.2}$$
$$\phi \otimes (\bar{\pi} \circ \pi) \otimes \psi - (\phi \otimes \pi \otimes {}_{\bar{\pi}}\psi), \tag{2.3}$$

where

$$\phi_\pi := \left( m \xrightarrow{\phi} m' \triangleleft c \xrightarrow{1_{m'} \triangleleft \pi} m' \triangleleft c' \right) \tag{2.4}$$
$${}_{\bar{\pi}}\psi := \left( \bar{c} \triangleright n \xrightarrow{\bar{\pi} \triangleright 1_n} \bar{\bar{c}} \triangleright n \xrightarrow{\psi} n' \right) \tag{2.5}$$

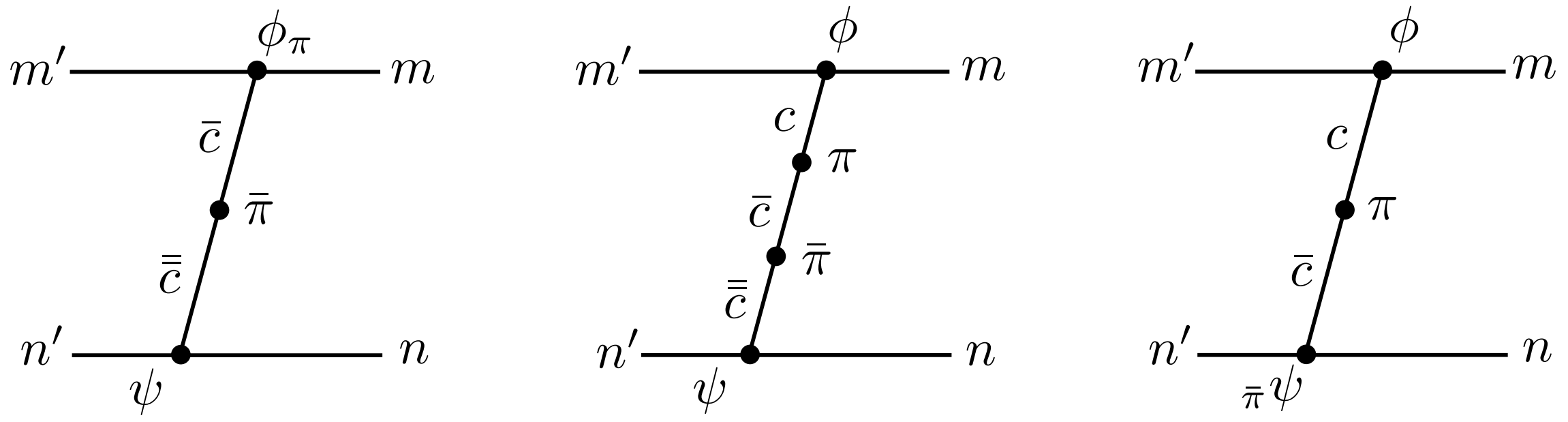

(The 1-morphisms are drawn from right to left, and the tensor products are drawn from top to bottom through this paper.)

The composition $(\phi' \otimes \pi' \otimes \psi') \circ (\phi \otimes \pi \otimes \psi)$ is defined to be $(\phi'' \otimes \pi'' \otimes \psi'')$ where

- $\pi'' \in \mathrm{Hom}_C(c' \otimes c, \bar{c}' \otimes \bar{c})$ is equal to

$$c' \otimes c \xrightarrow{\pi' \otimes \pi} \overline{c'} \otimes \bar{c},$$

- $\phi'' \in \mathrm{Hom}_M(m, m'' \triangleleft (c' \otimes c))$ is equal to

$$m \xrightarrow{\phi} m' \triangleleft c \xrightarrow{\phi' \triangleleft 1_c} (m'' \triangleleft c') \triangleleft c \underset{\sim}{\xrightarrow{\alpha}} m'' \triangleleft (c' \otimes c),$$

- $\psi'' \in \mathrm{Hom}_N((\bar{c}' \otimes \bar{c}) \triangleright n, n'')$ is equal to

$$(\bar{c}' \otimes \bar{c}) \triangleright n \underset{\sim}{\xrightarrow{\alpha}} \bar{c}' \triangleright (\bar{c} \triangleright n) \xrightarrow{1_{\bar{c}'} \triangleright \psi} \bar{c}' \triangleright n' \xrightarrow{\psi'} n''.$$

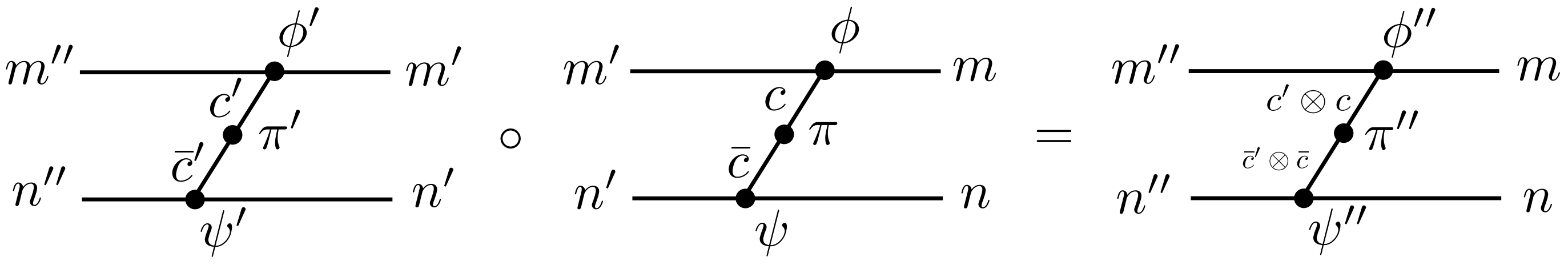

It is straightforward to check that this composition law respects the relations defining $\mathrm{Hom}_{M \otimes_C N}$. ◇

**Remark 2.6** We abuse notation by writing $(\phi \otimes \pi \otimes \psi)$ to denote the corresponding equivalence class in $\mathrm{Hom}_{M \otimes_C N}((m, n), (m', n'))$. We refer to a morphism $(\phi \otimes \pi \otimes \psi)$ as a skein. ◇

**Remark 2.7** In definition 2.1, we require C to have right duals. The construction makes sense even without this assumption. To clarify the role played by the duals, we define the general preskein category cor C not necessarily having duals at all in A.14 and A.15. ◇

**Remark 2.8** Using the defining relations, we see that

$$(\phi \otimes \pi \otimes \psi) = (\phi_\pi \otimes \mathrm{id}_{\bar{c}} \otimes \psi) = (\phi \otimes \mathrm{id}_c \otimes_\pi \psi).$$

◇

**Lemma 2.9** Any morphism in $M \otimes_C N$ can be written as a sum of composites of morphisms of the form $\phi \otimes \mathrm{id}_1 \otimes \psi : (m, n) \to (m', n')$ and $\mathrm{id}_{m \triangleleft c} \otimes \mathrm{id}_c \otimes \mathrm{id}_{c \triangleright n} : (m \triangleleft c, n) \to (m, c \triangleright n)$. ◇

*Proof.* First, any morphism is a sum of morphisms of the form $\phi \otimes \pi \otimes \psi$. By remark 2.8, we can take $\pi = \mathrm{id}_c$. Now $(\phi \otimes \mathrm{id}_c \otimes \psi) = (\mathrm{id}_{m'} \otimes \mathrm{id}_1 \otimes \psi) \circ (\mathrm{id}_{m' \triangleleft c} \otimes \mathrm{id}_c \otimes \mathrm{id}_{c \triangleright n}) \circ (\phi \otimes \mathrm{id}_1 \otimes \mathrm{id}_n)$. ∎

**Definition 2.10** . We define a functor $M \otimes N \to M \otimes_C N$ by the identity on objects, and $\phi \otimes \psi \mapsto \phi \otimes \mathrm{id}_1 \otimes \psi$ on morphisms. We define also morphisms $\beta_{m,c,n} : (m \triangleleft c, n) \to (m, c \triangleright n)$ by $\beta_{m,c,n} = \mathrm{id}_{m \triangleleft c} \otimes \mathrm{id}_c \otimes \mathrm{id}_{c \triangleleft n}$ .

$\beta_{m,c,n} =$

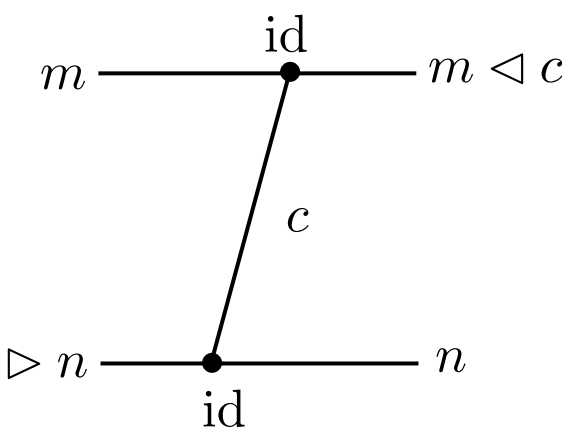

◇

**Lemma 2.11** The collection of morphisms $\beta_{m,c,n}$ defines a natural transformation

$$M \otimes C \otimes N \; \underset{\mathrm{id}_M \otimes \triangleright}{\overset{\triangleleft \otimes \mathrm{id}_N}{\rightrightarrows}} \; M \otimes_C N \qquad (\Downarrow \beta) .$$

$\diamond$

*Proof.* Given $\phi : m \to m'$, $\psi : n \to n'$ and $\pi : c \to \bar{c}$ we must check that the following diagram commutes.

$$\begin{array}{ccc} (m \triangleleft c, n) & \xrightarrow{\beta_{m,c,n}} & (m, c \triangleright n) \\ {\scriptstyle (\phi \triangleleft \pi) \otimes \mathrm{id}_1 \otimes \psi} \downarrow & & \downarrow {\scriptstyle \phi \otimes \mathrm{id}_1 \otimes (\pi \triangleright \psi)} \\ (m' \triangleleft \bar{c}, n') & \xrightarrow[\beta_{m',\bar{c},n'}]{} & (m', \bar{c} \triangleright n') \end{array}$$

Using the definition of composition and the relations in the definition of $\mathrm{Hom}_{M \otimes_C N}$ one can check that both composites equal $(\phi \triangleleft \mathrm{id}_c) \otimes \pi \otimes (\mathrm{id}_{\bar{c}} \triangleright \psi)$. ■

**Lemma 2.12** If every object in $C$ has a right dual, then the natural transformation $\beta$ described above is an isomorphism. $\diamond$

*Proof.* Given $(m, c, n)$ we provide an explicit inverse for $\beta_{m,c,n}$. Let $(c^*, \eta, \epsilon)$ denote a right dual for $c$. We define $\beta^{-1}_{m,c,n} = (\mathrm{id}_m \triangleleft \eta) \otimes \mathrm{id}_{c^*} \otimes (\epsilon \triangleright \mathrm{id}_n)$. The following is a proof that $\beta_{m,c,n} \circ \beta^{-1}_{m,c,n} = \mathrm{id}_{m,c \triangleright n}$.

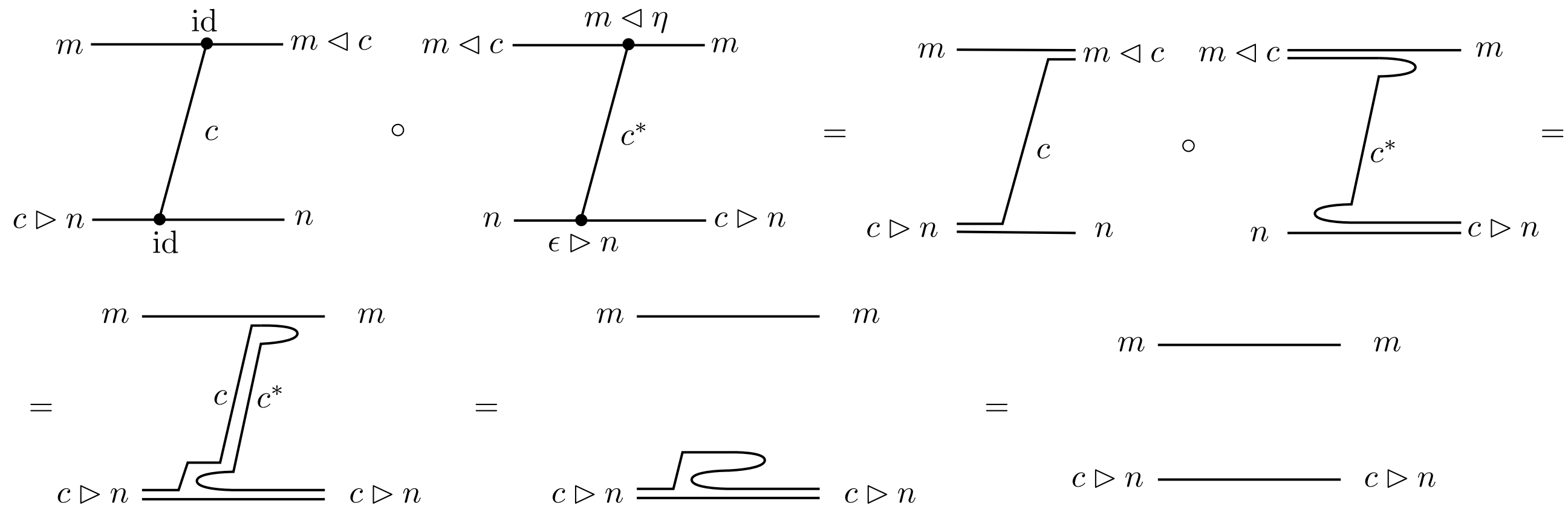

The proof that $\beta^{-1}_{m,c,n} \circ \beta_{m,c,n} = \mathrm{id}_{m \triangleleft c,n}$ is similar, using the same snake equation. ■

**Proposition 2.13** Suppose $C$ is a monoidal category with right duals and $M_C$, ${}_C N$ are $C$-module categories. Then 2.10 provides a $C$-balanced functor $M \otimes N \to M \otimes_C N$. $\diamond$

*Proof.* This is the content of lemmas 2.11 and 2.12. ■

Therefore, it is natural to assume $C$ to have right duals. We will do this frequently from now on. Recall that by definition any tensor category has right and left duals.

**Proposition 2.14** Suppose $C$ is a monoidal category with right duals and $M_C$, ${}_C N$ are $C$-module categories. Let $L$ be a category. Then composition with $M \otimes N \to M \otimes_C N$ induces an equivalence of categories $\mathrm{Fun}(M \otimes_C N, L) \to \mathrm{Fun}^{C-bal}(M \otimes N, L)$. $\diamond$

*Proof.* It follows from lemma 2.16, 2.18 and 2.19. ■

**Remark 2.15** Hence the pre-skein category $M \otimes_C N$ almost satisfies (and is characterized by) the universal property that defines the balanced tensor product $M \boxtimes_C N$, except that it is not abelian (thus not k-linear). We will fix this by defining the skein category $sk(M, C, N)$ as a certain completion of the pre-skein category. ◇

**Lemma 2.16** $\mathrm{Fun}(M \otimes_C N, L) \to \mathrm{Fun}^{C-bal}(M \otimes N, L)$ is surjective on objects. ◇

*Proof.* Given a C-balanced functor $F : M \otimes N \to L$ with balancing $\beta$, we will construct $\tilde{F} : M \otimes_C N \to L$ such that its composite with $M \otimes N \to M \otimes_C N$ is F. Define $\tilde{F} = F$ on objects. For morphisms, let $\phi \in \mathrm{Hom}_M(m, m' \lhd c)$, $\pi \in \mathrm{Hom}_C(c, \bar{c})$ and $\psi \in \mathrm{Hom}_N(\bar{c} \rhd n, n')$. We define $\tilde{F}([\phi \otimes \pi \otimes \psi])$ to be the composite

$$(m, n) \xrightarrow{F(\phi,n)} (m' \lhd c, n) \xrightarrow{F(m' \lhd \pi, n)} (m' \lhd \bar{c}, n) \xrightarrow{\beta_{m',\bar{c},n}} (m', \bar{c} \rhd n) \xrightarrow{F(m',\psi)} (m', n') \ .$$

The naturality of $\beta$ with respect to morphisms in C implies that this is equal to the composite

$$(m, n) \xrightarrow{F(\phi,n)} (m' \lhd c, n) \xrightarrow{\beta_{m',c,n}} (m', c \rhd n) \xrightarrow{F(m', \pi \rhd n)} (m', \bar{c} \rhd n) \xrightarrow{F(m',\psi)} (m', n') \ .$$

From these two expressions it is clear that this respects the relations in the definition of $\mathrm{Hom}_{M \otimes_C N}$. The proof that this respects composition is a straightforward calculation, using the naturality of $\beta$ with respect to morphisms in M and N. ∎

**Remark 2.17** Using lemma 2.9, the functor $\tilde{F} : M \otimes_C N \to L$ whose composite with $M \otimes N \to M \otimes_C N$ is the C-balanced functor $F : M \otimes N \to L$ with balancing $\beta$ is also determined by $\tilde{F}(m, n) = F(m, n)$, $\tilde{F}(\phi \otimes \mathrm{id}_1 \otimes \psi) = F(\phi, \psi)$ and $\tilde{F}(\mathrm{id}_{m \lhd c} \otimes \mathrm{id}_c \otimes \mathrm{id}_{c \rhd n}) = \beta_{m,c,n}$. ◇

**Lemma 2.18** $\mathrm{Fun}(M \otimes_C N, L) \to \mathrm{Fun}^{C-bal}(M \otimes N, L)$ is faithful. ◇

*Proof.* This is immediate from the fact that $M \otimes N \to M \otimes_C N$ is surjective on objects. ∎

**Lemma 2.19** $\mathrm{Fun}(M \otimes_C N, L) \to \mathrm{Fun}^{C-bal}(M \otimes N, L)$ is full. ◇

*Proof.* Given $\eta : F \to G$ in $\mathrm{Fun}^{C-bal}(M \otimes N, L)$ we must define $\tilde{\eta} : \tilde{F} \to \tilde{G}$ in $\mathrm{Fun}(M \otimes_C N, L)$. Denote by $\alpha$ and $\beta$ the balancing of F and G, respectively. Since $M \otimes N \to M \otimes_C N$ is surjective on objects, we are forced to define $\tilde{\eta}_{m,n} =: \eta_{m,n}$ for every object $(m, n) \in M \otimes_C N$. But we need to check that this defines a natural transformation. By lemma 2.9, it is enough to check that it is natural with respect to maps of the form $(\phi \otimes \mathrm{id}_1 \otimes \psi)$ and $(\mathrm{id}_{m \lhd c} \otimes \mathrm{id}_c \otimes \mathrm{id}_{c \rhd n})$. Naturality with respect to $(\phi \otimes \mathrm{id}_1 \otimes \psi)$ follows directly from the naturality of $\eta$. Now $\eta$ is balanced, which means that

$$\begin{array}{ccc} F(m \lhd c, n) & \xrightarrow{\eta_{m \lhd c, n}} & G(m \lhd c, n) \\ {\scriptstyle \alpha_{m,c,n}} \downarrow & & \downarrow {\scriptstyle \beta_{m,c,n}} \\ F(m, c \rhd n) & \xrightarrow[\eta_{m, c \rhd n}]{} & G(m, c \rhd n) \end{array}$$

commutes. Therefore

$$\begin{array}{ccc} \tilde{F}(m \lhd c, n) & \xrightarrow{\tilde{\eta}_{m \lhd c, n}} & \tilde{G}(m \lhd c, n) \\ {\scriptstyle \tilde{F}(\mathrm{id}_{m \lhd c} \otimes \mathrm{id}_c \otimes \mathrm{id}_{c \rhd n})} \downarrow & & \downarrow {\scriptstyle \tilde{G}(\mathrm{id}_{m \lhd c} \otimes \mathrm{id}_c \otimes \mathrm{id}_{c \rhd n})} \\ \tilde{F}(m, c \rhd n) & \xrightarrow[\tilde{\eta}_{m, c \rhd n}]{} & \tilde{G}(m, c \rhd n) \end{array}$$

commutes, i.e. $\eta$ is natural with respect to $(\mathrm{id}_{m \lhd c} \otimes \mathrm{id}_c \otimes \mathrm{id}_{c \rhd n})$. ∎

## 2.2 Skein Category sk(M,C,N)

**Definition 2.20** (Skein Category, $sk(M, C, N)$)
Let $C$ be a monoidal category with right duals, and let $M_C$, ${}_CN$ be module categories over $C$. Define the skein category $sk(M, C, N)$ to be the Cauchy completion $Cau(M \otimes_C N)$ (cf A.2, A.6) of the pre-skein category $M \otimes_C N$. ⋄

**Remark 2.21** There is a $C$-balanced functor $M \otimes N \to sk(M, C, N)$ obtained by the composite $M \otimes N \to M \otimes_C N \to sk(M, C, N)$, where the first arrow is the $C$-balanced functor in 2.13 (thus the composite is automatically $C$-balanced), and the second is the canonical functor to (A.2). ⋄

**Remark 2.22** An object in $sk(M, C, N)$ is an idempotent matrix $A : (m_1, n_1) \oplus \cdots \oplus (m_k, n_k) \to (m_1, n_1) \oplus \cdots \oplus (m_k, n_k)$ whose $(i, j)$-entry is a morphism $(m_j, n_j) \to (m_i, n_i)$ in $M \otimes_C N$. ⋄

**Proposition 2.23** Let $L$ be Cauchy complete and suppose $C$ has right duals. Then composition with the $C$-balanced functor $M \otimes N \to sk(M, C, N)$ induces an equivalence of categories

$$Fun(sk(M, C, N), L) \to Fun^{C-bal}(M \otimes N, L).$$

⋄

*Proof.* We wish to show the composite

$$Fun(sk(M, C, N), L) \to Fun(M \otimes_C N, L) \to Fun^{C-bal}(M \otimes N, L)$$

is an equivalence. The first functor is an equivalence by the universal property of the Cauchy completion (A.3) and the second one is an equivalence by proposition 2.14. ■

**Remark 2.24** Comparing 2.23 with 1.10, the defining universal property of balanced tensor product, we know we are closed to proving that the skein category $sk(M, C, N)$ is indeed equivalent to the balanced tensor product. What is missing so far are

1. $sk(M, C, N)$ is k-linear (in particular, abelian) (cf 2.30).
2. The equivalence in 2.23 has to be restricted to their right-exact counterparts: $Lin(M \boxtimes_C N, L) \to Bilin^{C-bal}(M \otimes N, L)$ (cf 2.31).

⋄

## 2.3 Proof of the Main Theorem

We will show that $M \otimes N \to sk(M, C, N)$ presents $sk(M, C, N)$ as the balanced tensor product $M \boxtimes_C N$ whenever $C$ is a finite tensor category and $M$, $N$ and $M \boxtimes_C N$ are finite semisimple. To quickly convince the reader that such result is correct, we also proved a degenerated statement $sk(M, C, C) \simeq M$ directly in A.62.

**Remark 2.25** In [DSS19], the existence of the balanced tensor product is established, when $C$ is a finite tensor category and $M_C, {}_C N$ are finite k-linear $C$-module categories. That is the reason for the appearance of the finiteness conditions on $M$, $C$, $N$ and the rigidity assumption on $C$ in the statements of all results in the present paper which involve the balanced tensor product. ⋄

In [DSS19] the authors remark that their construction of $M \boxtimes_C N$ allows them to extend its universal property (as described in 1.10) to the case when the target category $L$ is just a finitely cocomplete category. We record this in the following lemma.

**Lemma 2.26** Let $C$ be a finite tensor category, $M_C$, ${}_C N$ finite k-linear C-module categories, and $L$ a finitely cocomplete category. (Thus the balanced tensor product $M \boxtimes_C N$ exists by [DSS19].) Then the restriction functor is an equivalence:

$$\mathrm{Lin}(M \boxtimes_C N, L) \xrightarrow{\sim} \mathrm{Bilin}^{C-bal}(M \otimes N, L).$$

◇

*Proof.* If $L$ is assumed to be k-linear, this follows from the universal property of the balanced tensor product. Here, $L$ is only assumed to be finitely cocomplete, and a proof is outlined in [DSS19, Remark 3.4]. ■

**Notation 2.27** Given a category $M$, we denote by $\mathrm{Fin}(M)$ its finite cocompletion (A.11). ◇

Recall from 2.13 that we have a C-balanced functor $M \otimes N \to M \otimes_C N$ provided that $C$ has right duals. Composing it with $M \otimes_C N \to \mathrm{Fin}(M \otimes_C N)$ provides a C-balanced functor $M \otimes N \to \mathrm{Fin}(M \otimes_C N)$ (because the first one is C-balanced).

**Lemma 2.28** Let $C$ be a monoidal category having right duals, let $M_C$ and ${}_C N$ be module categories over $C$, and let $L$ be a finitely cocomplete category. Then the restriction map

$$\mathrm{Lin}(\mathrm{Fin}(M \otimes_C N), L) \to \mathrm{Fun}^{C-bal}(M \otimes N, L)$$

is an equivalence. ◇

*Proof.* That $\mathrm{Lin}(\mathrm{Fin}(M \otimes_C N), L) \simeq \mathrm{Fun}(M \otimes_C N, L)$ follows from a claim in A.11, and that $\mathrm{Fun}(M \otimes_C N, L) \simeq \mathrm{Fun}^{C-bal}(M \otimes N, L)$ follows from 2.14. ■

**Proposition 2.29** Suppose $C$ is a finite tensor category and $M_C$ and ${}_C N$ are k-linear finite semisimple module categories. Then we have a canonical equivalence of categories $\mathrm{Fin}(M \otimes_C N) \simeq M \boxtimes_C N$. ◇

*Proof.* Since $M, N$ are semisimple, any functor $M \otimes N \to L$ is bilinear (right exact in each variable). So by 2.28 the map

$$\mathrm{Lin}(\mathrm{Fin}(M \otimes_C N), L) \to \mathrm{Bilin}^{C-bal}(M \otimes N, L)$$

defined by composing with $M \otimes N \to \mathrm{Fin}(M \otimes_C N)$ is an equivalence for any finitely cocomplete $L$. By 2.26, composing with $M \otimes N \to M \boxtimes_C N$ gives an equivalence

$$\mathrm{Lin}(M \boxtimes_C N, L) \xrightarrow{\sim} \mathrm{Bilin}^{C-bal}(M \otimes N, L)$$

for any finitely cocomplete category $L$.

Thus, $\mathrm{Fin}(M \otimes_C N)$ and $M \boxtimes_C N$ are both finitely cocomplete categories and are both characterized by the same universal property. Therefore, they are canonically equivalent. ■

Now we prove that $\mathrm{sk}(M, C, N)$ is a finite semisimple k-linear category, whenever $M$, $N$ and $M \boxtimes_C N$ are finite semisimple k-linear categories.

**Lemma 2.30** Let $C$ be a finite tensor category, and let $M_C, {}_C N$ be k-linear finite semisimple C-module categories. Suppose $M \boxtimes_C N$ is also k-linear finite semisimple. Then $\mathrm{sk}(M, C, N)$ is k-linear finite semisimple.

◇

*Proof.* We have a fully faithful inclusion $\mathrm{sk}(M, C, N) = \mathrm{Cau}(M \otimes_C N) \hookrightarrow \mathrm{Fin}(M \otimes_C N)$. By proposition 2.29, we have $\mathrm{Fin}(M \otimes_C N) \simeq M \boxtimes_C N$, because $M$, $N$ are finite semisimple. Therefore $\mathrm{sk}(M, C, N)$ is finite semisimple, by proposition A.9. ■

**Lemma 2.31** Let $\mathcal{C}$ be a finite tensor category, and let $\mathcal{M}_{\mathcal{C}}, {}_{\mathcal{C}}\mathcal{N}$ be $k$-linear finite semisimple $\mathcal{C}$-module categories. Suppose $\mathcal{M} \boxtimes_{\mathcal{C}} \mathcal{N}$ is also $k$-linear finite semisimple. Then the $\mathcal{C}$-balanced functor $\mathcal{M} \otimes \mathcal{N} \to \mathrm{sk}(\mathcal{M}, \mathcal{C}, \mathcal{N})$ exhibits $\mathrm{sk}(\mathcal{M}, \mathcal{C}, \mathcal{N})$ as the balanced tensor product $\mathcal{M} \boxtimes_{\mathcal{C}} \mathcal{N}$. $\diamond$

*Proof.* Recall that proposition 2.23 asserts that the restriction map

$$\mathrm{Fun}(\mathrm{sk}(\mathcal{M}, \mathcal{C}, \mathcal{N}), \mathcal{L}) \xrightarrow{\sim} \mathrm{Fun}^{\mathcal{C}-bal}(\mathcal{M} \otimes \mathcal{N}, \mathcal{L})$$

is an equivalence for every Cauchy complete category $\mathcal{L}$. Comparing this with the defining universal property of balanced tensor product (cf 1.10), we need to show that $\mathrm{sk}(\mathcal{M}, \mathcal{C}, \mathcal{N})$ is $k$-linear (in particular, abelian), and to restrict the equivalence to their right-exact counterparts:

$$\mathrm{Lin}(\mathrm{sk}(\mathcal{M}, \mathcal{C}, \mathcal{N}), \mathcal{L}) \xrightarrow{\sim} \mathrm{Bilin}^{\mathcal{C}-bal}(\mathcal{M} \otimes \mathcal{N}, \mathcal{L}).$$

By 2.30 we know $\mathrm{sk}(\mathcal{M}, \mathcal{C}, \mathcal{N})$ is $k$-linear, so that resolves the first part, and we also know that it is semisimple, so

$$\mathrm{Fun}(\mathrm{sk}(\mathcal{M}, \mathcal{C}, \mathcal{N}), \mathcal{L}) = \mathrm{Lin}(\mathrm{sk}(\mathcal{M}, \mathcal{C}, \mathcal{N}), \mathcal{L}).$$

Finally,

$$\mathrm{Fun}^{\mathcal{C}-bal}(\mathcal{M} \otimes \mathcal{N}, \mathcal{L}) = \mathrm{Bilin}^{\mathcal{C}-bal}(\mathcal{M} \otimes \mathcal{N}, \mathcal{L})$$

because $\mathcal{M}$ and $\mathcal{N}$ are semisimple. ■

**Remark 2.32** Collecting results from [DSS21], we can conclude that $\mathcal{M} \boxtimes_{\mathcal{C}} \mathcal{N}$ is finite semisimple (so lemma 2.31 applies) in any of the following situations (in decreasing level of generality):

- $\mathcal{C}$ is a finite semisimple tensor category and ${}_{\mathrm{Vect}_k}\mathcal{M}_{\mathcal{C}}$, ${}_{\mathcal{C}}\mathcal{N}_{\mathrm{Vect}_k}$ are separable bimodule categories ([DSS21, proposition 2.5.3, theorem 2.5.5]);
- $k$ is perfect, $\mathcal{C}$ is a separable tensor category and $\mathcal{M}_{\mathcal{C}}, {}_{\mathcal{C}}\mathcal{N}$ are finite semisimple $\mathcal{C}$-module categories ([DSS21, proposition 2.5.10]);
- $k$ has characteristic $0$, $\mathcal{C}$ is a finite semisimple tensor category and $\mathcal{M}_{\mathcal{C}}, {}_{\mathcal{C}}\mathcal{N}$ are finite semisimple $\mathcal{C}$-module categories ([DSS21, Corollary 2.6.9]);

$\diamond$

Since we are mostly interested in the characteristic zero case, we record the following direct consequence of lemma 2.31.

**Theorem 2.33** Suppose $k$ has characteristic $0$, $\mathcal{C}$ is a finite semisimple tensor category, and $\mathcal{M}_{\mathcal{C}}, {}_{\mathcal{C}}\mathcal{N}$ are $k$-linear finite semisimple $\mathcal{C}$-module categories. Then the $\mathcal{C}$-balanced functor $\mathcal{M} \otimes \mathcal{N} \to \mathrm{sk}(\mathcal{M}, \mathcal{C}, \mathcal{N})$ exhibits $\mathrm{sk}(\mathcal{M}, \mathcal{C}, \mathcal{N})$ as the balanced tensor product $\mathcal{M} \boxtimes_{\mathcal{C}} \mathcal{N}$. $\diamond$

**Remark 2.34** One can immediately generalize 2.1 to obtain a definition of $\mathcal{M}^1 \otimes_{\mathcal{C}_1} \mathcal{M}^2 \otimes_{\mathcal{C}_2} \ldots \otimes_{\mathcal{C}_{n-1}} \mathcal{M}^n$ where morphisms are skeins as in the picture below.

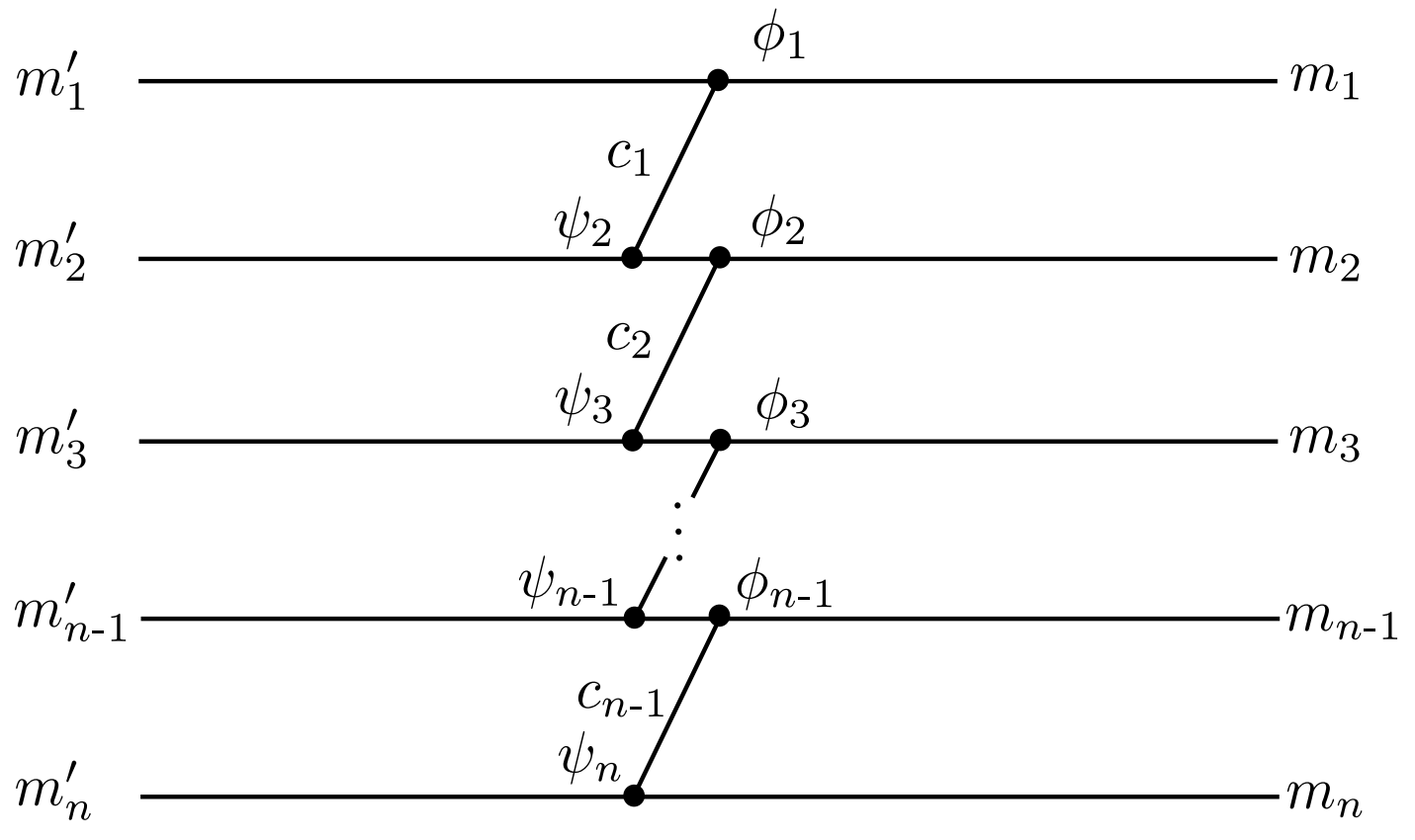

Then one can define $sk(M^1, C_1, M^2, C_2, \ldots, C_{n\text{-}1}, M^n)$ by Cauchy completion and thus extend Theorem 2.33 to iterated products.

$\diamond$

**Corollary 2.35** Suppose $k$ is a field of characteristic $0$. Let $C_1, \ldots, C_{n-1}$ be finite semisimple tensor categories, and let $M^1_{C_1}, {}_{C_1}M^2_{C_2}, \ldots, {}_{C_{n\text{-}1}}M^n$ be k-linear finite semisimple bimodule categories. Then we have an equivalence of categories

$$sk(M^1, C_1, M^2, C_2, \ldots, C_{n\text{-}1}, M^n) \simeq M^1 \boxtimes_{C_1} M^2 \boxtimes_{C_2} \ldots \boxtimes_{C_{n\text{-}1}} M^n.$$

$\diamond$

**Remark 2.36** Just like theorem 2.33, corollary 2.35 also holds in any of the more general settings described in remark 2.32. $\diamond$

# A Appendix

## A.1 Completions

In this subsection, we collect results about completions of categories.

**Definition A.1** (Cauchy Complete)
A ($\mathrm{Vect}_k$-enriched) category $M$ is Cauchy complete if it has all finite direct sums and all idempotents split.

$\diamond$

**Definition A.2** (Cauchy Completion (abstract version)) [Kel05, Sections 5.5 and 5.7].
Given a category $M$, we denote by $\mathrm{Cau}(M)$ the Cauchy completion of $M$, the smallest subcategory of $\mathrm{Fun}(M^{op}, \mathrm{Vect}_k)$ containing all the representables and closed under finite direct sums and retracts. Its objects can be identified as those $X \in \mathrm{Fun}(M^{op}, \mathrm{Vect}_k)$ such that

$$\mathrm{Hom}_{\mathrm{Fun}(M^{op}, \mathrm{Vect}_k)}(X, -) : \mathrm{Fun}(M^{op}, \mathrm{Vect}_k) \to \mathrm{Vect}_k$$

preserves all small colimits. Alternatively, they are the retracts of finite direct sums of representables. $\diamond$

**Lemma A.3** For any category $M$, the category $\mathrm{Cau}(M)$ is Cauchy complete. The Yoneda embedding induces a fully faithful functor $M \hookrightarrow \mathrm{Cau}(M)$, restriction of which gives an equivalence:

$$\mathrm{Fun}(\mathrm{Cau}(M), L) \xrightarrow{\sim} \mathrm{Fun}(M, L)$$

for any Cauchy complete category $L$ $\diamond$

*Proof.* See [Kel05, Sections 5.5 and 5.7]. ■

Next, we recall the usual explicit construction of the Cauchy completion $\mathrm{Cau}(M)$ of a ($\mathrm{Vect}_k$-enriched) category $M$ in A.6.

**Definition A.4** (Matrix Category)
Given a category $M$, we define $\mathrm{Mat}(M)$ to be the category whose objects are tuples of objects in $M$. We denote such an object by $m_1 \oplus \cdots \oplus m_k$. We then define

$$\mathrm{Hom}_{\mathrm{Mat}(M)}(m_1 \oplus \cdots \oplus m_k, n_1 \oplus \cdots \oplus n_\ell)$$

to be the k-vector space of $(\ell \times k)$ matrices whose $(i, j)$-th entry is a morphism $m_j \to n_i$ in $M$. Composition is defined by matrix product. $\diamond$

**Definition A.5** (Karoubi Completion)
Given a category $M$, we define its Karoubi completion $Kar(M)$ to be the category whose objects are idempotent endomorphisms $m \xrightarrow{p} m$ in $M$, and the hom space $\mathrm{Hom}_{Kar(C)}(m \xrightarrow{p} m, n \xrightarrow{q} n)$ is the $k$-linear subspace of $\mathrm{Hom}_M(m, n)$ consisting of $f$ such that the following diagram commutes.

$$\begin{array}{ccc} m & \xrightarrow{f} & n \\ \downarrow p & \searrow^{f} & \downarrow q \\ m & \xrightarrow[f]{} & n \end{array}$$

◇

**Definition A.6** (Cauchy Completion (explicit version))
The explicit construction of the Cauchy completion is given by $Cau(M) = Kar(Mat(M))$. ◇

So an object in $Cau(M)$ is an idempotent matrix

$$m_1 \oplus \cdots \oplus m_k \xrightarrow{A} m_1 \oplus \cdots \oplus m_k.$$

We introduce a few lemmas about Cauchy complete categories.

**Lemma A.7** Suppose we have a fully faithful functor $F : M \hookrightarrow N$ where $M$ is Cauchy complete. If $F(x) = a' \oplus b'$ then we have $x = a \oplus b$ and isomorphisms $F(a) \simeq a'$, $F(b) \simeq b'$ such that the following diagrams commute.

$$\begin{array}{ccccc} F(a) & \longrightarrow & F(x) & \longleftarrow & F(b) \\ \simeq\downarrow & & \downarrow = & & \downarrow\simeq \\ a' & \longrightarrow & F(x) & \longleftarrow & b' \end{array} \qquad \begin{array}{ccccc} F(a) & \longleftarrow & F(x) & \longrightarrow & F(b) \\ \simeq\downarrow & & \downarrow = & & \downarrow\simeq \\ a' & \longleftarrow & F(x) & \longrightarrow & b' \end{array}$$

◇

*Proof.* Denote by $i_{a'} : a' \to F(x)$, $r_{a'} : F(x) \to a'$ and $p_{a'} : F(x) \to F(x)$ the corresponding inclusion, retraction and idempotent, so that $r_{a'}i_{a'} = \mathrm{id}_{a'}$ and $i_{a'}r_{a'} = p_{a'}$, and similarly for $b'$. We have additional equations $r_{a'}i_{b'} = 0 = r_{b'}i_{a'}$ and $p_{a'} + p_{b'} = \mathrm{id}_{F(x)}$. Now $F$ is full, so there exists $p_a : x \to x$ such that $F(p_a) = p_{a'}$ and similarly for $p_b$. Now $M$ is idempotent complete, so $p_a$ splits, i.e. we obtain $i_a : a \to x$ and $r_a : x \to a$ such that $r_a i_a = \mathrm{id}_a$ and $i_a r_a = p_a$ and similarly for $b$. Now $F(p_a + p_b) = p_{a'} + p_{b'} = \mathrm{id}_{F(x)}$. Since $F$ is faithful we get $p_a + p_b = \mathrm{id}_x$. This means that $x = a \oplus b$. Finally $r_{a'}F(i_a) : F(a) \to a'$ and $r_{b'}F(i_b) : F(b) \to b'$ are the desired isomorphisms. ∎

**Lemma A.8** Suppose we have a fully faithful functor $F : M \hookrightarrow N$ where $M$ is Cauchy complete and $N$ is $k$-linear. Suppose every short exact sequence splits in $N$. Then $M$ is $k$-linear and every short exact sequence splits in $M$. ◇

*Proof.* We start by showing that $M$ is $k$-linear (essentially, we need to check that $M$ is abelian). We now show that $M$ has kernels, and they are preserved by $F$. (The proof for cokernels is dual to this proof).

So let $x \to y$ be a morphism in $M$. Then $F(x) \to F(y)$ has a kernel $k' \to F(x)$ in $N$. The short exact sequence $0 \to k' \to F(x) \to c' \to 0$ splits, where $c'$ is the cokernel of $k' \to F(x)$. So we get $F(x) = k' \oplus c'$. Then, by lemma A.7, we have $x = k \oplus c$ where $F(k) \simeq k'$ and

$$\begin{array}{ccc} F(k) & \longrightarrow & F(x) \\ \simeq\downarrow & & \downarrow = \\ k' & \longrightarrow & F(x) \end{array}$$

commutes. This means $F(k) \to F(x)$ is also a kernel of $F(x) \to F(y)$. But $F$ is fully faithful, so it reflects limits, hence $k \to x$ is a kernel of $x \to y$.

Next, we show that every monomorphism is a kernel. (The proof that every epimorphism is a cokernel is dual to this proof). Let $a \hookrightarrow x$ be a monomorphism in $M$. Since $F$ preserves kernels, we know $F(a) \hookrightarrow F(x)$ is a monomorphism. Moreover, if $x \to c$ is the cokernel of $a \hookrightarrow x$, then $F(x) \to F(c)$ is the cokernel of $F(a) \hookrightarrow F(x)$ (because $F$ also preserves cokernels). So $0 \to F(a) \to F(x) \to F(c) \to 0$ is exact, so $F(x) = F(a) \oplus F(c)$. Then $x = a \oplus c$ (by lemma A.7) and so $a \to x$ is the kernel of $x \to c$.

This concludes the proof that $M$ is k-linear, and that $F$ is exact.

Finally, we need that $F : M \to N$ is exact. The fact that every short exact sequence splits in $M$ follows easily from lemma A.7, and the facts that $F$ is exact and every short exact sequence splits in $N$. ■

**Proposition A.9** Suppose we have a fully faithful functor $M \hookrightarrow N$, where $M$ is Cauchy complete and $N$ is k-linear finite semisimple. Then $M$ is a k-linear finite semisimple category. ◇

*Proof.* By lemma A.8 the category $M$ also k-linear and every short exact sequence splits in $M$, which implies that every object in $M$ is projective and also that $F : M \to N$ exact. Since $F$ is faithful and $N$ is finite, all hom spaces in $M$ are finite dimensional.

Since $F$ is exact and fully faithful, it preserves and reflects monomorphisms, therefore $x \in M$ is simple if and only if $F(x) \in N$ is simple. So in particular, given $N$ has finitely many isomorphism classes of simple objects, the same is true of $M$.

If $F(x) = \bigoplus_{i=1}^{n} x_i'$ with $x_i'$ simple, then by induction and lemma A.7 we have $x = \bigoplus_{i=1}^{n} x_i$ where $F(x_i) \simeq x_i'$ so $x_i$ is simple. ■

**Definition A.10** (Finitely Cocomplete)
A category $M$ is finitely cocomplete if it has all finite colimits. ◇

**Definition A.11** (Finite Cocompletion, $\mathrm{Fin}(M)$) [Kel05, Section 5.7], [Lóp13, Section 2.2.1].
Given a category $M$, we denote by $\mathrm{Fin}(M)$ its finite cocompletion. It is the smallest finitely cocomplete subcategory of $\mathrm{Fun}(M^{op}, \mathrm{Vect}_k)$ that contains all the representables. Its objects can be identified as those $X \in \mathrm{Fun}(M^{op}, \mathrm{Vect}_k)$ such that $\mathrm{Hom}_{\mathrm{Fun}(M^{op},\mathrm{Vect}_k)}(X, -) : \mathrm{Fun}(M^{op}, \mathrm{Vect}_k) \to \mathrm{Vect}_k$ preserves filtered colimits. Alternatively, they are the coequalisers of pairs of morphisms between finite coproducts of representables. The Yoneda embedding induces a fully faithful functor $M \hookrightarrow \mathrm{Fin}(M)$. ◇

**Lemma A.12** For any finitely cocomplete $L$, the restriction map $\mathrm{Lin}(\mathrm{Fin}(M), L) \to \mathrm{Fun}(M, L)$ is an equivalence. ◇

*Proof.* See [Kel05, Section 5.7] or [Lóp13, Section 2.2.1]. ■

**Remark A.13** From our descriptions of $\mathrm{Cau}(M)$ (A.2) and $\mathrm{Fin}(M)$ (A.11), it is immediate that we have fully faithful functors $M \hookrightarrow \mathrm{Cau}(M) \hookrightarrow \mathrm{Fin}(M)$. ◇

## A.2 Rigidity

This section explains the role of duals in the theory of the balanced Kelly tensor product (of finitely cocomplete categories). Namely, the existence of left or right duals in $C$ allows for the category $M \otimes_C N$ to have finite dimensional hom spaces, whenever $M$, $N$ and $C$ have finite dimensional hom spaces. The relevance of having both left and right duals (rigidity) remains unclear to us from this perspective, although rigidity is a standing assumption in the most general known construction of the balanced Deligne tensor product of finite k-linear categories (see [DSS19]).

Similar to the preskein category $M \otimes_C N$, we define the general preskein category $M \hat{\otimes}_C N$ that works for monoidal categories $C$ not necessarily with duals in A.14, A.15. We then prove that they are in fact isomorphic categories when $C$ has weak duals (A.18).

**Definition A.14** (General Preskein Category, $\mathcal{M}\hat{\otimes}_{\mathcal{C}}\mathcal{N}$)
Let $\mathcal{C}$ be a monoidal category (not necessarily with duals) and let $\mathcal{M}_{\mathcal{C}}$ and ${}_{\mathcal{C}}\mathcal{N}$ be $\mathcal{C}$-module categories. We define $\mathcal{M}\hat{\otimes}_{\mathcal{C}}\mathcal{N}$ as the category whose objects are pairs $(m, n)$ and whose morphism spaces are the spaces of alternating skeins of arbitrary (finite) length, as depicted below.

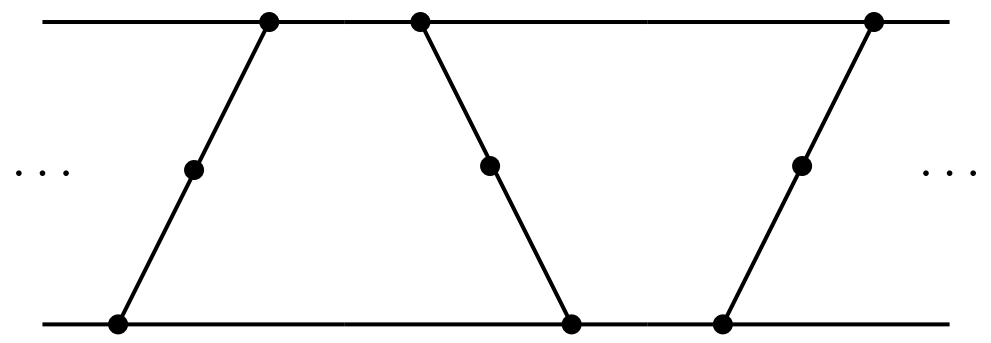

This should of course be interpreted as a suitable direct sum of tensor products of morphism spaces in $\mathcal{M}$, $\mathcal{C}$ and $\mathcal{N}$. We impose relations analogous to those in definition 2.1, plus the extra relations below.

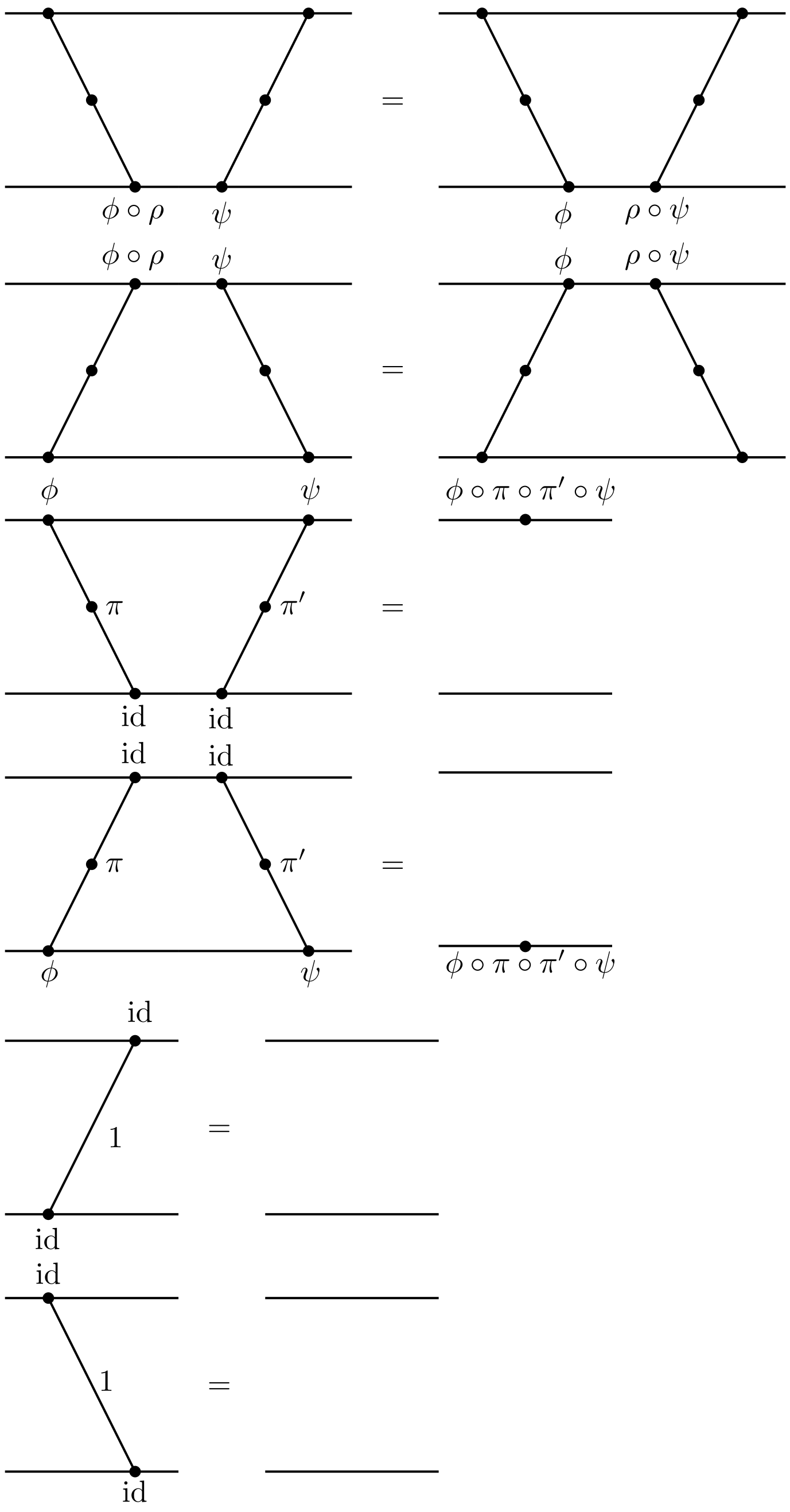

Each of these relations may be applied locally to any finite length alternating skein. In the case of the last two relations, after applying the relation one must apply the composition rule from definition 2.1 to obtain an alternating skein again.

Composition is defined by horizontally stacking the two diagrams and then applying the composition rule from definition 2.1 if there are parallel diagonal wires. $\diamond$

**Definition A.15** (General Preskein Category (alternative))
Let $C$ be a monoidal category and let $M_C$ and ${}_CN$ be $C$-module categories. We define $M\hat{\otimes}_C N$ by starting with $M \otimes N$, then freely adjoining morphisms $(m \lhd c, n) \to (m, c \rhd n)$ and $(m, c \rhd n) \to (m \lhd c, n)$ for each $m \in M$, $n \in N$ and $c \in C$ (depicted below)

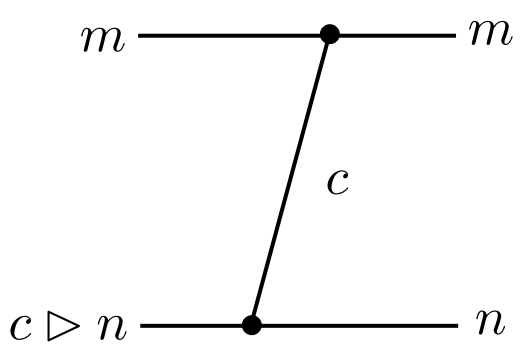

$: (m \lhd c, n) \to (m, c \rhd n)$

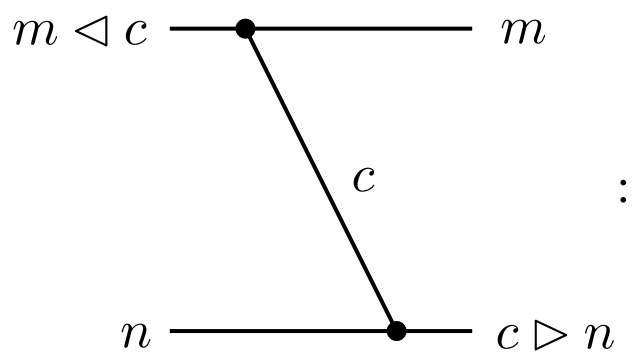

$: (m, c \rhd n) \to (m \lhd c, n)$

and finally imposing the relations below (where we denote composition by concatenation).

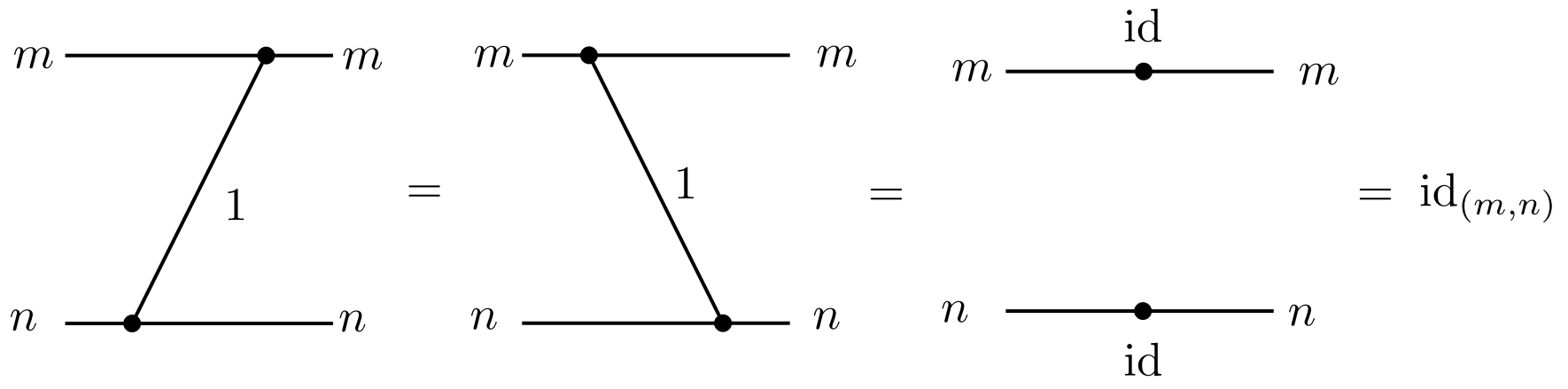

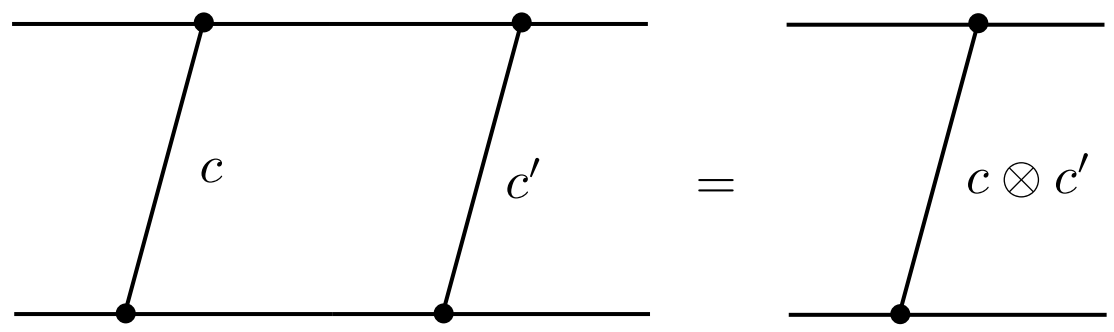

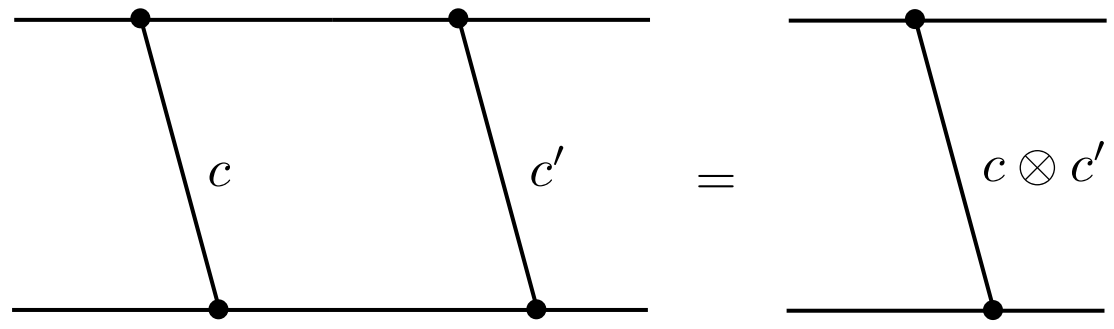

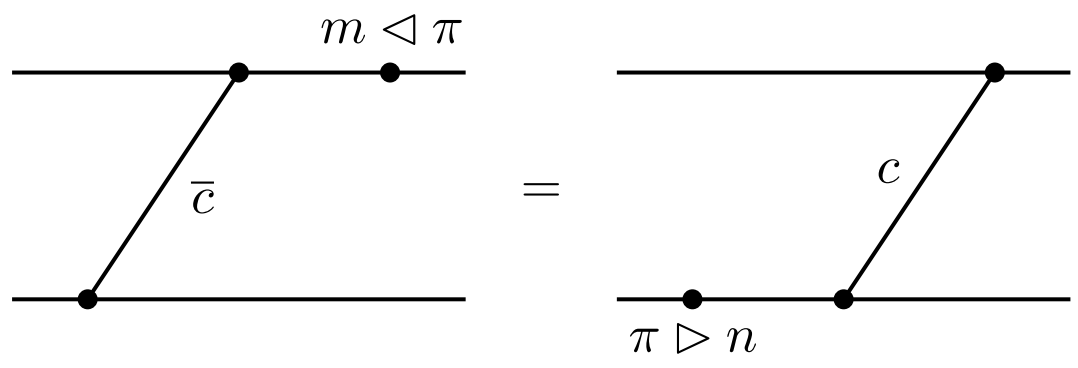

where $\pi : c \to \bar{c}$ is a morphism in $C$

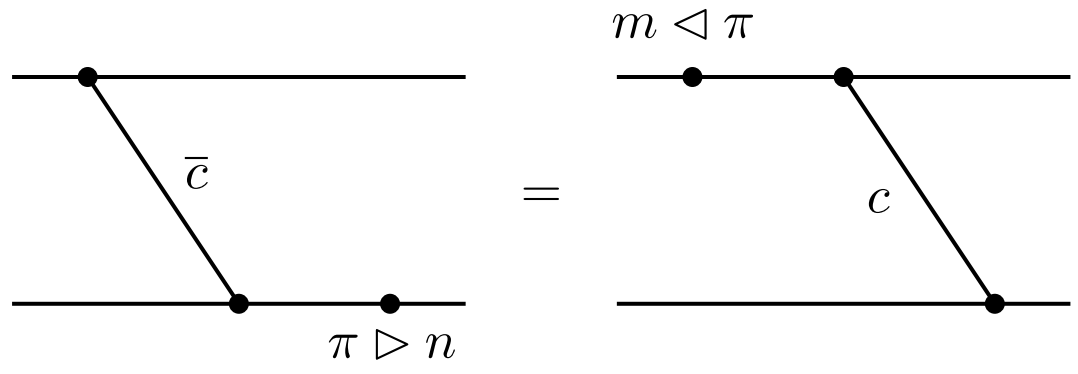

where $\pi : c \to \bar{c}$ is a morphism in $C$

◇

**Definition A.16** Given an object $c$ in a monoidal category $C$, a weak right dual for $c$ is a triple $(c^*, \eta, \epsilon)$, where $c^*$ is an object in $C$ and $\eta : 1 \to c \otimes c^*$ and $\epsilon : c^* \otimes c \to 1$ are morphisms in $c$, such that the following equation holds.

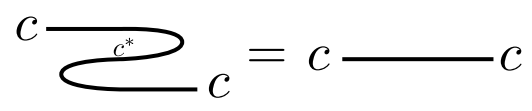

A monoidal category with weak right duals is a monoidal category where every object has a weak right dual. ◇

**Lemma A.17** Let $C$ be a monoidal category with weak right duals and let $M_C$ and ${}_CN$ be $C$-module categories. Suppose $(c^*, \eta, \epsilon)$ is a weak right dual for $c$. Then

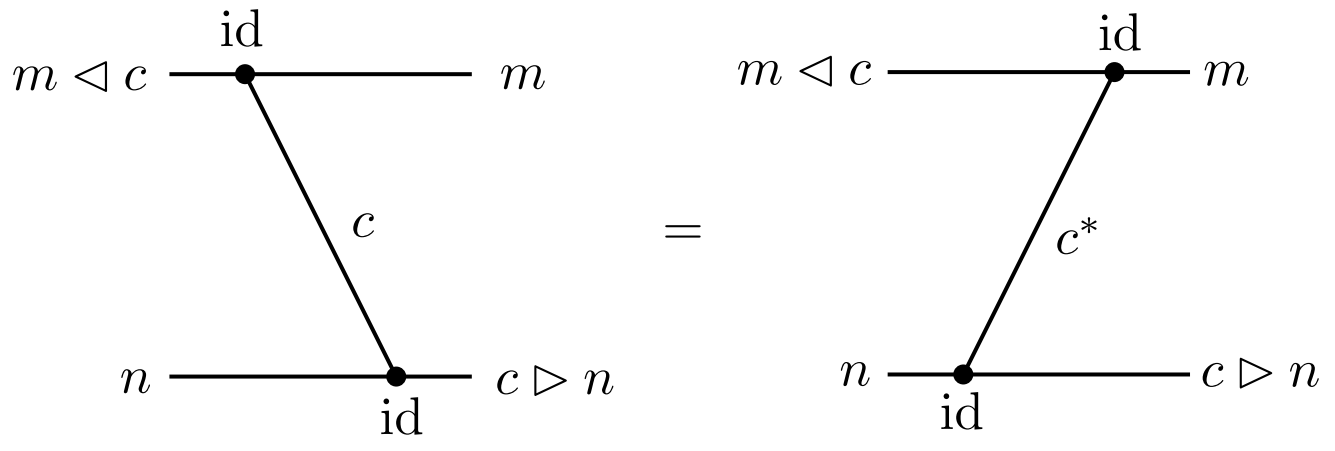

in $M\hat{\otimes}_C N$. ◇

*Proof.* Both sides of the equation are inverse to

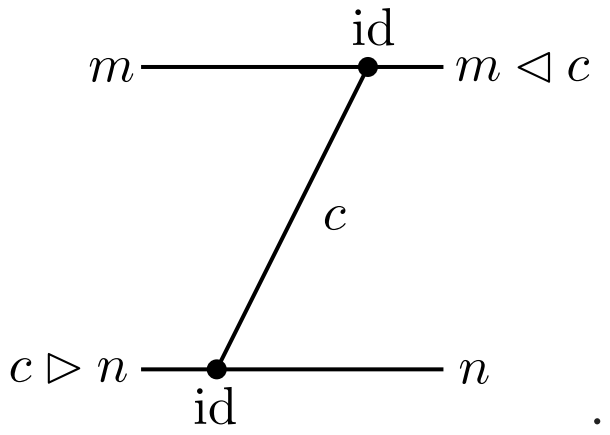

For the left hand side, this follows from the relations in definition A.14. For the right hand side, it follows from the proof of lemma 2.12, where only the fact that $(c^*, \eta, \epsilon)$ is a weak right dual for $c$ is used. ■

**Lemma A.18** Let $C$ be a monoidal category with weak right duals and let $M_C$ and ${}_CN$ be $C$-module categories. Then the preskein category $M \otimes_C N$ and the general preskein category $M\hat{\otimes}_C N$ are isomorphic. ◇

*Proof.* There is an obvious functor $M \otimes_C N \to M\hat{\otimes}_C N$ which is a bijection on objects. To show that it is an isomorphism on morphism spaces, we use lemma A.17 to reduce every morphism in $M\hat{\otimes}_C N$ to a morphism in $M \otimes_C N$. ■

**Proposition A.19** Let $C$ be a monoidal category, let $M_C$ and ${}_CN$ be $C$-module categories and let $L$ be a category. Then composition with $M \otimes N \to M \otimes_C N$ induces an equivalence of categories $\mathrm{Fun}(M\hat{\otimes}_C N, L) \to \mathrm{Fun}^{C-\mathrm{bal}}(M \otimes N, L)$. ◇

*Proof.* The only difference between this and the proof of proposition 2.14 is that now providing an inverse to the balancing $\beta_{m,c,n} : (m \triangleleft c, n) \to (m, c \triangleright n)$ (see lemma 2.12) does not require duals, as we can simply use the skein with a $c$ wire going up. ■

**Remark A.20** While read from right to left, the wires in the definition of preskein category go down, while the wires in the definition of the general preskein category go both ways. We could also define the "opposite" preskein category where the diagonal wires go up (read from right to left). This category is isomorphic to the general preskein category $M\hat{\otimes}_C N$ of definition A.14 when $C$ has weak left duals (defined in an analogous way to weak right duals). When $C$ has both weak left and right duals (in particular, when $C$ is rigid), then all three categories are isomorphic. ◇

## A.3 The Non-semisimple Case

In this subsection, we extend the construction of the skein category $\mathrm{sk}(M, C, N)$ to the non-semisimple skein category $M \boxtimes^{sk}_C N$ for $C, M, N$ that are not necessarily semisimple. While this generalizes what we did in the main text of the paper, we put it in the appendix because the tools we use here rely on several results in [Kel05]. For convenience, we include the necessary results from [Kel05] in section A.3.1)

The general idea is to start from $M \otimes_C N$ and then add finite colimits, to obtain a finitely cocomplete category $M \boxtimes^{sk}_C N$. However, we also want the functor $M \otimes N \to M \boxtimes^{sk}_C N$ to be right exact in each variable.

When $M$, $N$ are semisimple, this is automatic, so we can simply take $M \boxtimes^{sk}_C N = \mathrm{Fin}(M \otimes_C N)$, i.e. the finite cocompletion of $M \otimes_C N$, i.e. the closure of the image of

$$M \otimes_C N \hookrightarrow \mathrm{Fun}((M \otimes_C N)^{op}, \mathrm{Vect}_k)$$

under finite colimits.

When $M$, $N$ are not necessarily semisimple, we must simultaneously complete $M \otimes_C N$ under finite colimits and also force diagrams coming from colimit diagrams in $M, N$ to become colimits in $M \boxtimes^{sk}_C N$. This is achieved by changing from $\mathrm{Fun}$ to $\mathrm{Bilex}$ (both are the same in the semisimple case). More precisely, we define $M \boxtimes^{sk}_C N$ as the closure under finite colimits of the image of

$$M \otimes_C N \hookrightarrow \mathrm{Fun}((M \otimes_C N)^{op}, \mathrm{Vect}_k) \xrightarrow{R} \mathrm{Bilex}((M \otimes_C N)^{op}, \mathrm{Vect}_k),$$

where, the first functor is the Yoneda embedding, and the second functor $R$ is right adjoint to the inclusion $\mathrm{Bilex}((M \otimes_C N)^{op}, \mathrm{Vect}_k) \hookrightarrow \mathrm{Fun}((M \otimes_C N)^{op}, \mathrm{Vect}_k)$ of the full subcategory. The existence of $R$ is the most substantial result from [Kel05] that we rely on.

**Definition A.21** (Kelly Balanced Tensor Product)
Suppose $C$ is a monoidal category and $M_C, {}_C N$ are finitely cocomplete $C$-module categories. Their balanced Kelly tensor product is a finitely cocomplete category $M \boxtimes^K_C N$ together with a $C$-balanced bilinear functor $M \otimes N \to M \boxtimes^K_C N$ such that precomposition defines an equivalence

$$\mathrm{Lin}(M \boxtimes^K_C N, L) \to \mathrm{Bilin}^{C-bal}(M \otimes N, L)$$

for any finitely cocomplete category $L$. ◇

Suppose $C$ is a monoidal category and $M_C, {}_C N$ are finitely cocomplete $C$-module categories. We explain how to construct a category $M \boxtimes^{sk}_C N$ so that $M \otimes N \to M \boxtimes^{sk}_C N$ presents it as the balanced Kelly tensor product of $M$ and $N$ over $C$. We follow very closely the procedure adopted in [Lóp13] to construct the unbalanced Kelly tensor product, starting from $M \otimes N$.

**Definition A.22** A functor $(M \otimes_C N)^{op} \to L$ is said to be left exact in each variable if the composite $M^{op} \otimes N^{op} \to (M \otimes_C N)^{op} \to L$ is left exact in each variable. In other words it sends finite colimits in $M$ and finite colimits in $N$ to finite limits in $L$. We denote by $\mathrm{Bilex}((M \otimes_C N)^{op}, \mathrm{Vect}_k) \subset \mathrm{Fun}(M \otimes_C N)^{op}, \mathrm{Vect}_k)$ the full subcategory whose objects are those functors which are left exact in each variable. ◇

**Definition A.23** We say that a functor $M \otimes_C N \to L$ is bilinear if the composite $M \otimes N \to M \otimes_C N \to L$ is bilinear (i.e. right exact in each variable). We denote by $\mathrm{Bilin}(M \otimes_C N, L) \subset \mathrm{Fun}(M \otimes_C N, L)$ the full subcategory whose objects are the bilinear functors. ◇

The subcategory $\mathrm{Bilex}((M \otimes_C N)^{op}, \mathrm{Vect}_k) \subset \mathrm{Fun}((M \otimes_C N)^{op}, \mathrm{Vect}_k)$ is reflective (see [Kel05, theorem 6.5] or theorem A.43) so that we have an adjunction

$$\mathrm{Bilex}((M \otimes_C N)^{op}, \mathrm{Vect}_k) \underset{i}{\overset{R}{\underset{\hookrightarrow}{\overset{\leftarrow}{\perp}}}} \mathrm{Fun}((M \otimes_C N)^{op}, \mathrm{Vect}_k) .$$

In particular, $\mathrm{Bilex}((M \otimes_C N)^{op}, \mathrm{Vect}_k)$ is cocomplete.

**Definition A.24** We denote by $K : M \otimes_C N \to \mathrm{Bilex}((M \otimes_C N)^{op}$ the composite

$$M \otimes_C N \xrightarrow{Y} \mathrm{Fun}((M \otimes_C N)^{op}, \mathrm{Vect}_k) \xrightarrow{R} \mathrm{Bilex}((M \otimes_C N)^{op}, \mathrm{Vect}_k)\ ,$$

where $Y$ is the Yoneda embedding. ◇

**Lemma A.25** The functor $K : M \otimes_C N \to \mathrm{Bilex}((M \otimes_C N)^{op}$ is bilinear. ◇

*Proof.* This follows from assertion (5.51) in [Kel05]. But we can give the following proof. We must show that $\mathrm{Bilex}((M \otimes_C N)^{op}, \mathrm{Vect}_k)(RY-, \phi) : (M \otimes_C N)^{op} \to \mathrm{Vect}_k$ is left exact in each variable, for any $\phi \in \mathrm{Bilex}((M \otimes_C N)^{op}, \mathrm{Vect}_k)$. This follows from the calculation

$$\mathrm{Bilex}((M \otimes_C N)^{op}, \mathrm{Vect}_k)(RY-, \phi) \simeq \mathrm{Fun}((M \otimes_C N)^{op}, \mathrm{Vect}_k)(Y-, \phi) \simeq \phi$$

where the first step uses the adjunction $R \dashv i$ and the second step is the Yoneda lemma. ■

**Definition A.26** Let $C$ be a monoidal category and let $M_C$, ${}_CN$ be finitely cocomplete $C$-module categories. We define $M \boxtimes^{sk}_C N$ as the closure of the full image of the functor

$$K : M \otimes_C N \to \mathrm{Bilex}((M \otimes_C N)^{op}, \mathrm{Vect}_k)$$

under finite colimits. This means $M \boxtimes^{sk}_C N$ is the smallest replete full subcategory of $\mathrm{Bilex}((M \otimes_C N)^{op}, \mathrm{Vect}_k)$ which contains every object of the form $K(m, n)$ and is closed under finite colimits. Thus $M \boxtimes^{sk}_C N$ is finitely cocomplete and comes equipped with a functor $Z : M \otimes_C N \to M \boxtimes^{sk}_C N$. ◇

**Lemma A.27** The functor $Z : M \otimes_C N \to M \boxtimes^{sk}_C N$ is bilinear. ◇

*Proof.* This follows from lemma A.25 and the fact that the fully faithful functor

$$M \boxtimes^{sk}_C N \hookrightarrow \mathrm{Bilex}((M \otimes_C N)^{op}, \mathrm{Vect}_k)$$

reflects colimits. ■

**Proposition A.28** Given a finitely cocomplete category $L$, the functor

$$\mathrm{Lin}(M \boxtimes^{sk}_C N, L) \to \mathrm{Bilin}(M \otimes_C N, L)$$

given by precomposition with $Z$ is an equivalence. ◇

*Proof.* See [Kel05, theorem 6.23] or theorem A.59. ■

**Lemma A.29** The functor $\mathrm{Bilin}(M \otimes_C N, L) \to \mathrm{Bilin}^{C-bal}(M \otimes N, L)$ is an equivalence, for any category $L$. ◇

*Proof.* By lemma 2.14 we know that $\mathrm{Fun}(M \otimes_C N, K) \to \mathrm{Fun}^{C-bal}(M \otimes N, L)$ is an equivalence. By definition, a functor $M \otimes_C N \to L$ is bilinear if and only if the composite $M \otimes N \to M \otimes_C N \to L$ is bilinear, so we obtain an equivalence between the two subcategories. ■

**Theorem A.30** Let $C$ be a monoidal category and let $M_C$, ${}_CN$ be finitely cocomplete module categories. Then the $C$-balanced functor $M \otimes N \to M \boxtimes^{sk}_C N$ presents $M \boxtimes^{sk}_C N$ as the balanced Kelly tensor product of $M$ and $N$ over $C$. ◇

*Proof.* This follows from proposition A.28 and lemma A.29. ■

**Corollary A.31** Let $C$ be a finite tensor category, $M_C$, ${}_CN$ finite k-linear $C$-module categories. Then $M \boxtimes_C^{sk} N$ is k-linear and the $C$-balanced functor $M \otimes N \to M \boxtimes_C^{sk} N$ presents $M \boxtimes_C^{sk} N$ as the balanced Deligne tensor product of $M$ and $N$ over $C$. ◇

*Proof.* This follows from theorem A.30 and lemma 2.26. ■

The following lemma makes the connection between the construction of $M \boxtimes_C^{sk} N$ in this section and our previous constructions in the semisimple case.

**Lemma A.32** Let $C$ be a monoidal category, and let $M_C$ and ${}_CN$ be k-linear semisimple $C$-module categories. Then $M \boxtimes_C^{sk} N \simeq \mathrm{Fin}(M \otimes_C N)$. ◇

*Proof.* In this case we have $\mathrm{Bilex}((M \otimes_C N)^{op}, \mathrm{Vect}_k) = \mathrm{Fun}((M \otimes_C N)^{op}, \mathrm{Vect}_k)$. So the functor $K$ in definition A.26 is simply the Yoneda embedding $M \otimes_C N \hookrightarrow \mathrm{Fun}((M \otimes_C N)^{op}, \mathrm{Vect}_k)$ and so $M \boxtimes_C^{sk} N$ is simply the finite cocompletion of $M \otimes_C N$. ■

**Corollary A.33** Let $C$ be a monoidal category, and let $M_C$ and ${}_CN$ be k-linear semisimple $C$-module categories. Then $M \otimes_C N \to \mathrm{Fin}(M \otimes_C N)$ presents $\mathrm{Fin}(M \otimes_C N)$ as the balanced Kelly tensor product of $M$ and $N$ over $C$. ◇

**Remark A.34** In lemma A.32 and corollary , semisimplicity of $M$ and $N$ can be replaced by the weaker condition that every short exact sequence splits in $M$ and $N$. ◇

### A.3.1 Review of some basic concepts of enriched category theory

Here we review the concepts from [Kel05] that are needed in A.3.

First we note that the results in [Kel05] can be applied to categories enriched over very general kinds of symmetric monoidal categories $V$. To be precise, the results we need apply when $V$ is a locally bounded, closed symmetric monoidal category. We won't need to define these terms, as we won't prove the results from [Kel05]. All that needs to be said is that $\mathrm{Vect}_k$ satisfies these properties, and we will always take $V = \mathrm{Vect}_k$.

**Definition A.35** Consider a functor $F : I \to A$. A cone over $F$ is an object $a \in A$ together with a natural transformation $a \Rightarrow F$ (where $a$ denotes the constant functor $a : I \to A$). A cocone under $F$ is an object $b \in A$ together with a natural transformation $F \Rightarrow b$. ◇

**Definition A.36** Let $A$ be a category. A cone in $A$ consists of a category $I$, a functor $F : I \to A$ and a cone over $F$. A cocone in $A$ consists of a category $I$, a functor $F : I \to A$ and a cocone under $F$. ◇

We often denote a cocone in $A$ simply by writing $a_i \to a$ which denotes $a_i = F(i)$ and the component $a_i \to a$ of the natural transformation corresponding to $i \in I$.

**Definition A.37** A small sketch is a pair $(A^{op}, \Phi)$ where $A$ is a small category and $\Phi$ is a small set of cocones in $A$. We often refer to a cocone in $\Phi$ as a $\Phi$-cocone in $A$. ◇

**Remark A.38** Note that in [Kel05] $\Phi$ is allowed to be a set of cylinders. Cocones are particular kinds of cylinders and we won't need the extra generality. ◇

**Definition A.39** Given a small sketch $(A^{op}, \Phi)$ and a category $B$, a $\Phi$-comodel in $B$ is a functor $A \to B$ sending the $\Phi$-cocones in $A$ to colimit cocones in $B$. We denote by $\Phi - \mathrm{Com}(A, B) \subset \mathrm{Fun}(A, B)$ the full subcategory whose objects are the $\Phi$-comodels. A $\Phi$-algebra is a functor $A^{op} \to \mathrm{Vect}_k$ sending all $\Phi$-cocones in $A$ (which become cones in $A^{op}$) to limit cones in $\mathrm{Vect}_k$. We denote by $\Phi - \mathrm{Alg} \subset \mathrm{Fun}(A^{op}, \mathrm{Vect}_k)$ the full subcategory whose objects are the $\Phi - algebras$. ◇

**Definition A.40** Given a small sketch $(A^{op}, \Phi)$, we define $\Theta_\Phi$ to be the set of all morphisms in $\mathrm{Fun}(A^{op}, \mathrm{Vect}_k)$ of the form $\mathrm{colim}_i Y(a_i) \to Y(a)$ where $a_i \to a$ is a $\Phi$-cocone in $A$ and $Y$ is the Yoneda embedding. ◇

**Definition A.41** Given a category $P$, we say that an object $p \in P$ is orthogonal to a morphism $\theta : m \to n$ in $P$ if $\theta^* : \mathrm{Hom}_P(n, p) \to \mathrm{Hom}_P(m, p)$ is an isomorphism. ◇

**Definition A.42** A full subcategory $C \subset P$ is called reflective if the fully faithful inclusion functor $C \hookrightarrow P$ has a left adjoint $R : P \to C$. The left adjoint is called the reflector. ◇

**Theorem A.43** [Kel05, theorem 6.5]
Let $A$ be a small category and let $\Theta$ be a set of morphisms in $\mathrm{Fun}(A^{op}, \mathrm{Vect}_k)$. Let $C_\Theta \subset \mathrm{Fun}(A^{op}, \mathrm{Vect}_k)$ be the full subcategory whose objects are those orthogonal to every $\theta \in \Theta$. Then $C_\Theta$ is a reflective subcategory.
◇

**Remark A.44** The category $C_\Theta$ is denoted $\Theta$-Alg in [Kel05]. ◇

**Lemma A.45** [Kel05, theorem 6.11] Let $(A^{op}, \Phi)$ be a small sketch. Then the reflective subcategory $C_{\Theta_\Phi} \subset \mathrm{Fun}(A^{op}, \mathrm{Vect}_k)$ is the category $\Phi$-Alg. ◇

*Proof.* An object $\phi \in \mathrm{Fun}(A^{op}, \mathrm{Vect}_k)$ is orthogonal to a morphism $\mathrm{colim}_i Y(a_i) \to Y(a)$ in $\Theta_\Phi$ if

$$\mathrm{Fun}(A^{op}, \mathrm{Vect}_k)(Y(a), \phi) \to \mathrm{Fun}(A^{op}, \mathrm{Vect}_k)(\mathrm{colim}_i Y(a_i), \phi)$$

is an isomorphism. That is to say $\phi(a) \to \lim_i \phi(a_i)$ is an isomorphism, i.e. $\phi(a) \to \phi(a_i)$ is a limit cone. So the objects of $\mathrm{Fun}(A^{op}, \mathrm{Vect}_k)$ which are orthogonal to every $\theta \in \Theta_\Phi$ are exactly the $\Phi$-algebras. ■

**Definition A.46** Given a small sketch $(A^{op}, \Phi)$ we denote by $C_\Phi$ the category $\Phi - \mathrm{Alg}$. ◇

So we obtain an adjunction

$$C_\Phi \underset{i_\Phi}{\overset{R_\Phi}{\underset{\hookrightarrow}{\overset{\longleftarrow}{\perp}}}} \mathrm{Fun}(A^{op}, \mathrm{Vect}_k) \ .$$

**Definition A.47** Given a small sketch $(A^{op}, \Phi)$, and denoting by $Y : A \to \mathrm{Fun}(A^{op}, \mathrm{Vect}_k)$ the Yoneda embedding, we define $K_\Phi : A \to C_\Phi$ as the composite functor

$$A \xhookrightarrow{Y} \mathrm{Fun}(A^{op}, \mathrm{Vect}_k) \xrightarrow{R_\Phi} C_\Phi \ .$$

◇

**Lemma A.48** Given a small sketch $(A^{op}, \Phi)$, the functor $K_\Phi : A \to C_\Phi$ is a $\Phi$-comodel, i.e. it sends $\Phi$-cocones in $A$ to colimit cocones in $C_\Phi$. ◇

*Proof.* This is (a special case of) statement (5.51) in [Kel05]. Since the proof is short, we can record it here. We must show that, given any object $\phi \in C_\Phi = \Phi - \mathrm{Alg}$, the functor $C_\Phi(R_\Phi Y-, \phi) : A^{op} \to \mathrm{Vect}_k$ is a $\Phi$-algebra, i.e. sends $\Phi$-cocones in $A$ to limit cones in $\mathrm{Vect}_k$. But $C_\Phi(R_\Phi Y-, \phi) \simeq \mathrm{Fun}(A^{op}, \mathrm{Vect}_k)(Y-, \phi) \simeq \phi$. ■

**Remark A.49** Since $C_\Phi \subset \mathrm{Fun}(A^{op}, \mathrm{Vect}_k)$ is a reflective subcategory and $\mathrm{Fun}(A^{op}, \mathrm{Vect}_k)$ is cocomplete, $C_\Phi$ is also cocomplete. ◇

**Definition A.50** A diagram type is just a small category. Given a diagram type $I$, a diagram of type $I$ in $D$ is a functor $I \to D$. Let $\mathcal{F}$ be a set of diagram types. An $\mathcal{F}$-colimit in a category $D$ is a colimit of a diagram of type $I$ where $I \in \mathcal{F}$. A functor $D \to B$ is $\mathcal{F}$-cocontinuous if it sends $\mathcal{F}$-colimits in $D$ to $\mathcal{F}$-colimits in $B$. We denote by $\mathcal{F} - \mathrm{Cocts}(D, B) \subset \mathrm{Fun}(D, B)$ the full subcategory whose objects are the $\mathcal{F}$-cocontinuous functors. ◇

**Remark A.51** In [Kel05] a more general notion of indexing type is used. Here we only need the more restricted notion of diagram type defined above. ◇

**Definition A.52** A subcategory $D \subset C$ is replete if given $d \in D$ and an isomorphism $f : d \to c$ in $C$, then $c$ and $f$ are also in $D$. ◇

**Definition A.53** Let $\mathcal{F}$ be a set of diagram types and suppose $C$ is $\mathcal{F}$-cocomplete. Let $K \subset C$ be a full subcategory. The closure of $K$ under $\mathcal{F}$-colimits in $C$ is the smallest replete full subcategory of $C$ containing $K$ and closed under $\mathcal{F}$-colimits. ◇

**Definition A.54** The type of a cocone $a \Rightarrow F$ is the type of the diagram $F$. ◇

**Definition A.55** Let $\mathcal{F}$ be a small set of diagram types. A small $\mathcal{F}$-sketch is a small sketch $(A^{op}, \Phi)$ where the type of every $\Phi$-cocone is in $\mathcal{F}$. ◇

**Definition A.56** Let $\mathcal{F}$ be a small set of diagram types and let $(A^{op}, \Phi)$ be a small $\mathcal{F}$-sketch. Define $D_{\Phi,\mathcal{F}}$ as the closure of the full image of $K_\Phi$ in $C_\Phi$ under $\mathcal{F}$-colimits. Notice that $K_\Phi : A \to C_\Phi$ factors through $D_{\Phi,\mathcal{F}} \hookrightarrow C_\Phi$, so we obtain a functor $Z_{\Phi,\mathcal{F}} : A \to D_{\Phi,\mathcal{F}}$. ◇

**Lemma A.57** Let $\mathcal{F}$ be a small set of diagram types and let $(A^{op}, \Phi)$ be a small $\mathcal{F}$-sketch. Then the functor $Z_{\Phi,\mathcal{F}} : A \to D_\Phi$ is a $\Phi$-comodel, i.e. it sends $\Phi$-cocones in $A$ to colimit cocones in $D_{\Phi,\mathcal{F}}$. ◇

*Proof.* This follows from the fact that $K_\Phi$ is a $\Phi$-comodel and the fully faithful inclusion $D_{\Phi,\mathcal{F}} \hookrightarrow C_\Phi$ reflects colimits. ■

**Lemma A.58** Let $\mathcal{F}$ be a small set of diagram types and let $(A^{op}, \Phi)$ be a small $\mathcal{F}$-sketch. Given an $\mathcal{F}$-cocontinuous functor $H : D_{\Phi,\mathcal{F}} \to B$, the composite $A \xrightarrow{Z_{\Phi,\mathcal{F}}} D_{\Phi,\mathcal{F}} \xrightarrow{H} B$ is a $\Phi$-comodel. ◇

*Proof.* Recall that $Z_{\Phi,\mathcal{F}}$ is a $\Phi$-comodel, i.e. it sends $\Phi$-cocones in $A$ to colimit cocones in $D_{\Phi,\mathcal{F}}$. Since $\mathcal{F}$ contains the type of every $\Phi$-cocone, $Z_{\Phi,\mathcal{F}}$ sends $\Phi$-cocones in $A$ to $\mathcal{F}$-colimit cocones in $D_{\Phi,\mathcal{F}}$. But $H$ sends $\mathcal{F}$-colimits in $D_{\Phi,\mathcal{F}}$ to $\mathcal{F}$-colimits in $B$, so finally the composite sends $\Phi$-cocones in $A$ to colimit cocones in $B$. ■

This means that precomposition with $Z_{\Phi,\mathcal{F}}$ defines a functor $Z^*_{\Phi,\mathcal{F}} : \mathcal{F}-\mathrm{Cocts}(D_{\Phi,\mathcal{F}}, B) \to \Phi-\mathrm{Com}(A, B)$ for any category $B$.

**Theorem A.59** [Kel05, theorem 6.23]
Let $\mathcal{F}$ be a small set of diagram types and let $(A^{op}, \Phi)$ be a small $\mathcal{F}$-sketch. Let $B$ be an $\mathcal{F}$-cocomplete category. Then

$$Z^*_{\Phi,\mathcal{F}} : \mathcal{F} - \mathrm{Cocts}(D_{\Phi,\mathcal{F}}, B) \to \Phi - \mathrm{Com}(A, B)$$

is an equivalence. ◇

In section A.3, we use the procedure described here to define $M \boxtimes^{sk}_C N$. We have $A = M \otimes_C N$, which is a small category whenever $M, N, C$ are small. We take $\mathcal{F}$ to be a set of finite categories containing exactly one representative of each isomorphism class (it is a small set). So $\mathcal{F}$-colimits are just finite colimits. We take $\Phi$ to be the set of all cocones of the kinds $m_i \otimes n \to m \otimes n$ and $m \otimes n_i \to m \otimes n$ where $m_i \to m$ is any finite colimit cocone in $M$ (with type in $\mathcal{F}$) and $n_i \to n$ is any finite colimit cocone in $N$ (with type in $\mathcal{F}$). The fact that we require the types to be in the small set $\mathcal{F}$ and the fact that the categories $M, N$ are small together imply that $\Phi$ is a small set. Then one obtains $C_\Phi = \Phi - Alg = \mathrm{Bilex}((M \otimes_C N)^{op}, \mathrm{Vect}_k)$ and $K_\Phi : M \otimes_C N \to \mathrm{Bilex}((M \otimes_C N)^{op}, \mathrm{Vect}_k)$ is the functor $K$ of definition A.24. Finally, the category $D_{\Phi,\mathcal{F}}$ obtained as the closure of the full image of $K_\Phi$ in $C_\Phi$ under finite colimits is exactly $M \boxtimes^{sk}_C N$ (definition A.26). Now $\mathcal{F}$-cocomplete is synonymous with finitely cocomplete and $\mathcal{F}$-cocontinuous means right exact. So the $\mathcal{F}$-cocontinuous functors $M \boxtimes^{sk}_C N \to L$ are exactly the linear functors. Moreover, $\Phi$-comodels $M \otimes_C N \to L$ are exactly the bilinear functors. So we obtain proposition A.28.

### A.3.2 Questions

We record here two questions which remain to be explored.

**Question A.60** Suppose $C$ is a monoidal category and $M_C$, ${}_CN$ are module categories (all enriched over $\mathrm{Vect}_k$). Under what conditions is the functor $M \otimes N \to M \otimes_C N$ right exact in each variable? $\diamond$

This is true for example in the unbalanced case, i.e. when $C = \mathrm{Vect}_k$. When this holds, one can show that the Yoneda embedding $M \otimes_C N \hookrightarrow \mathrm{Fun}((M \otimes_C N)^{op}, \mathrm{Vect}_k)$ factors through the inclusion $\mathrm{Bilex}((M \otimes_C N)^{op}, \mathrm{Vect}_k) \hookrightarrow \mathrm{Fun}((M \otimes_C N)^{op}, \mathrm{Vect}_k)$, yielding a functor $K : M \otimes_C N \hookrightarrow \mathrm{Bilex}((M \otimes_C N)^{op}, \mathrm{Vect}_k)$, so we can simply define $M \boxtimes_C^{sk} N$ as the closure of the image of this functor under finite colimits. This means we don't need to use the reflection $R : \mathrm{Fun}((M \otimes_C N)^{op}, \mathrm{Vect}_k) \to \mathrm{Bilex}((M \otimes_C N)^{op}, \mathrm{Vect}_k)$.

We note that in A.3 we have not made the assumption that the action of $C$ on $M$ and $N$ is right exact in each variable. It seems like this hypothesis should play a role in answering the question above.

**Question A.61** If the balanced tensor product of $k$-linear categories $M \boxtimes_C N$ exists, then must it agree with the Kelly tensor product $M \boxtimes_C^K N$ (and therefore with $M \boxtimes_C^{sk} N$)? $\diamond$

We know this is true when $C$ is a finite tensor category and $M$, $N$ are finite $k$-linear $C$-module categories (see [DSS19, Remark 3.4].) This also holds in the unbalanced case, i.e. when $C = \mathrm{Vect}_k$ (see [Lóp13, Theorem 18]).

## A.4 Misc. Results

This subsection gathers results, proofs, and arguments that would disrupt the flow of the main text. The content is unstructured, so readers should approach it as a collection of standalone items.
From the main theorem, we must have $\mathrm{sk}(M, C, C) \simeq M$. To quickly convince the reader that the main theorem is true before it is proven, we provide another direct proof for this equivalence:

**Proposition A.62** Let $C$ be a tensor category. Let $M_C$ be a right $C$-module category. Then we have an equivalence of categories $\mathrm{sk}(M, C, C) \simeq M$. $\diamond$

*Proof.* We provide two proofs. The first proof is immediate upon using the main theorem 2.33, and the known fact that $M \boxtimes_C C \simeq M$. The second proof is direct, without using the main theorem:

Construct the functor $M \to M \otimes C \to \mathrm{sk}(M, C, C)$, where the first functor is $m \mapsto (m, 1)$ and the second one is the canonical map (2.21). We contend that this is an equivalence of categories. It is straightforward to see that it is indeed fully faithful, so it suffices to show that it is essentially surjective. A typical object in the codomain $\mathrm{sk}(M, C, C)$ is some idempotent skein $I_c^m(\phi, \psi \,|\, d)$ (without loss of generality, assume $m$ to be simple). We contend that this object is isomorphic to $I_1^{m \triangleleft c}\left(\mu_{\phi,\psi}, 1 \,\middle|\, 1\right)$, where

$$\mu_{\phi,\psi} := m \triangleleft c \xrightarrow{\phi \triangleleft c} (m \triangleleft d) \triangleleft c \underset{\sim}{\xrightarrow{\alpha}} m \triangleleft (d \otimes c) \xrightarrow{m \triangleleft \psi} m \triangleleft c.$$

Indeed, the isomorphism is provided by the following two morphisms

$$I_1^{m \triangleleft c}\left(\mu_{\phi,\psi}, 1 \,\middle|\, 1\right) \xleftarrow{I_1^{m \triangleleft c}(\mu_{\phi,\psi}, 1 \,|\, 1) \circ {}_1^{mc} I_c^m(u, n \,|\, c^\star) \circ I_c^m(\phi, \psi \,|\, \overline{c})} I_c^m(\phi, \psi \,|\, \overline{c})$$
$$I_c^m(\phi, \psi \,|\, \overline{c}) \xleftarrow{I_c^m(\phi, \psi \,|\, \overline{c}) \circ {}_c^m I_1^{mc}(1, 1 \,|\, c) \circ I_1^{m \triangleleft c}(\mu_{\phi,\psi}, 1 \,|\, 1)} I_1^{m \triangleleft c}\left(\mu_{\phi,\psi}, 1 \,\middle|\, 1\right),$$

where $c^\star$ denotes the right dual of $c$, $n$ denotes the counit, and $u$ denotes the unit for $c$. The two given morphisms may seem unnecessarily long, but they have to be so by the definition of of Karoubi completion (in which objects must be "absorbed" into morphism). The following pictures show that they compose to the identities.

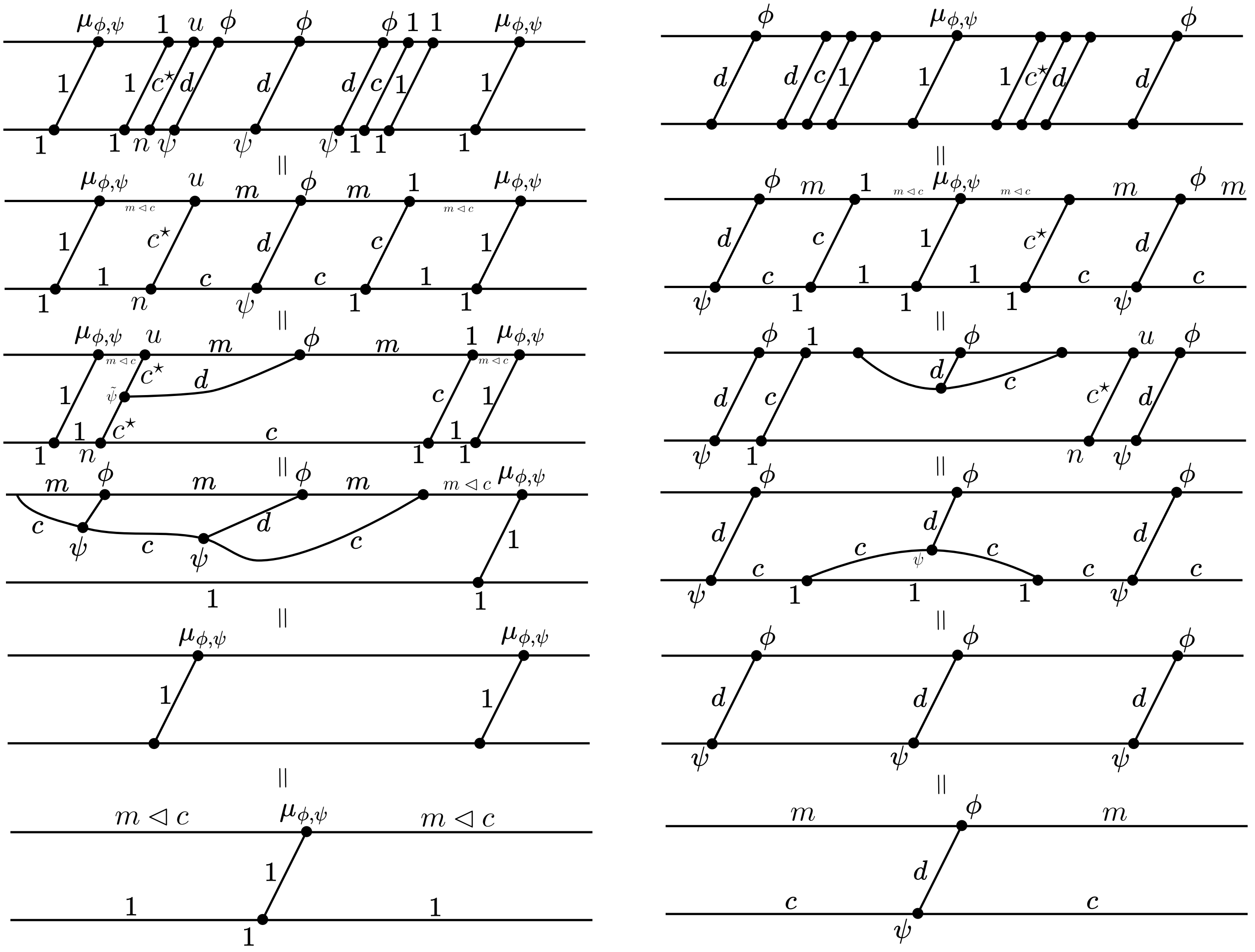

■